\documentclass[12pt]{article}

\usepackage{graphicx}
\usepackage{amsmath}
\usepackage{tabularx}
\usepackage{lscape}

\newcommand{\pr}[1]{\mathcal{P}^{(#1)}}

\begin{document}

\title{{\bf Universality of the baryon axial vector current operator in large-$N_c$ chiral perturbation theory}}

\author{Rub\'en Flores-Mendieta \\
	{\it \normalsize Instituto de F{\'\i}sica, Universidad Aut\'onoma de San Luis Potos{\'\i}} \\
	{\it \normalsize \'Alvaro Obreg\'on 64, Zona Centro, San Luis Potos{\'\i}, 78000, San Luis Potos{\'\i}, M\'exico}
        \and
        Guillermo S\'anchez-Almanza \\
	{\it \normalsize Instituto de F{\'\i}sica, Universidad Aut\'onoma de San Luis Potos{\'\i}} \\
	{\it \normalsize \'Alvaro Obreg\'on 64, Zona Centro, San Luis Potos{\'\i}, 78000, San Luis Potos{\'\i}, M\'exico}
	}

\maketitle

\begin{abstract}
The baryon axial vector current is computed in a combined formalism in $1/N_c$ and chiral corrections. Flavor $SU(3)$ symmetry breaking is accounted for in two ways: Implicitly through the integrals occurring in the one-loop graphs and explicitly through perturbative symmetry breaking. Loop integrals can be expanded in a power series in the ratio of the decuplet-octet baryon mass difference to the pseudoscalar meson mass and the first three terms in the series are retained and evaluated. The universal baryon axial vector current so constructed is neither diagonal nor off-diagonal in the sense that it can connect baryon states of either different or equal spins to obtain appropriate axial vector couplings. Processes of interest are found in octet-baryon and decuplet-baryon semileptonic decays and strong decays of decuplet baryons. A fit to the available experimental information is performed to determine the free parameters in the formalism, which allows one to estimate, for instance, the leading axial vector coupling in the semileptonic decays $\Omega^- \to \Xi^0 \ell^-\overline{\nu}_\ell$ and $\Omega^- \to {\Xi^*}^0\ell^-\overline{\nu}_\ell$.
\end{abstract}

\section{Introduction}

In 1991, Jenkins and Manohar proposed a formulation of chiral perturbation theory based on a chiral Lagrangian for broken $SU(3)_L\times SU(3)_R$ chiral symmetry, in which baryons are considered as static heavy fermions \cite{jm255,jm259,jen91}. The novelty in the formulation was the existence of an expansion in powers of momentum and light quark masses because the baryon mass did not appear explicitly in the Lagrangian. Furthermore, the chiral expansion in the strange quark mass required the involvement of both spin-1/2 and spin-3/2 baryons as explicit degrees of freedom. The success of heavy baryon chiral perturbation theory (HBChPT) was immediately evidenced in the computation of leading nonanalytic corrections to the baryon axial vector current produced by one-loop Feynman graphs. The chiral corrections were found to be large including only contributions from loop graphs involving octet field propagators \cite{jm255}. However, the inclusion of loop graphs with intermediate octet and decuplet baryon states gave rise to large cancellations among graphs \cite{jm259}. Later, it was discovered that these cancellations originate as a consequence of the emergent $SU(6)$ spin-flavor symmetry present in the $N_c \to \infty$ limit \cite{tHooft,ven,witten,dm1,djm94,djm95}, where $N_c$ is the number of colors, so it was found that the sum of all one-loop diagrams was of order $\mathcal{O}(1/N_c)$.

A combined framework in the chiral and $1/N_c$ expansions, hereafter loosely referred to as the combined formalism, is a computational scheme which has been used to extract low-energy consequences of QCD \cite{djm94}. The $1/N_c$ chiral Lagrangian in this framework was proposed in Ref.~\cite{jen96}. Shortly thereafter, a number of static properties of baryons have been dealt with. Among them baryon mass splittings \cite{djm94,jen96,weise}, baryon axial vector current \cite{djm94,rfm00,rfm06,rfm12,fg0,rfm21}, baryon vector current \cite{fmg,fg}, and baryon magnetic moment \cite{luty,rfm09,rfm14b,rfm21b,dai} stand out.

The research program developed within the combined formalism for the specific case of the baryon axial vector current, spanned over the past three decades, has been profitable. Early analyses found that large-$N_c$ consistency conditions imply that the leading nonanalytic corrections to the axial currents decrease as $1/N_c$ instead of increasing as $N_c$, as a naive analysis would indicate \cite{djm94}. Most importantly, it was found that this suppression only occurs if loop graphs are evaluated including the complete large-$N_c$ tower of intermediate states and axial couplings are used with ratios determined consistently in large $N_c$. Later work set the groundwork for a calculational scheme that simultaneously exhibits both the $m_q$ and $1/N_c$ expansions putting emphasis on how to deal with loop corrections in HBChPT so as to include the full dependence on the decuplet-octet baryon mass difference $\Delta$, while at the same time including the cancellations imposed by large-$N_c$ spin-flavor symmetry of baryons \cite{rfm00}. More recent works were devoted to present explicit calculations about the renormalization of the baryon axial vector current in different scenarios: The degeneracy limit $\Delta \to 0$ \cite{rfm06}, simultaneous effects of nonvanishing $\Delta$ and perturbative symmetry breaking (PSB) \cite{rfm12}, and the degeneracy limit again, but this time the baryon operators at $N_c=3$ involved in one-loop corrections are included in full \cite{rfm21}. The latter is particularly important because it allowed us to carry out a full comparison with the corresponding HBChPT results of Refs.~\cite{jm255,jm259}; the comparison was successful so it was concluded that the combined formalism and HBChPT yield the same results for $N_c=3$.

The objective of the present paper is to perform a calculation of the $SU(3)$ breaking corrections for axial vector coupling for different channels to $\mathcal{O}(p^2)$ in the chiral expansion and consistent with the strictures demanded by the $1/N_c$ expansion, extending the analysis of Ref.~\cite{rfm21} to evaluate nonvanishing $\Delta$ effects. These effects are accounted for by expanding loop integrals in powers of $\Delta/m$, where $m$ is the pseudoscalar meson mass. Physically interesting conclusions can be drawn by keeping the first three terms in the power expansion. The axial vector current operator so obtained is a universal one in the sense that its matrix elements between different baryon states (spin-1/2 and spin-3/2 baryons) yield axial vector coupling for different scenarios.

Nowadays, when comparing theoretical predictions with data, flavor $SU(3)$ breaking corrections to the static properties of baryons
cannot be ignored. In particular, SB corrections to the axial-vector couplings involving octet and decuplet baryons constitute an important issue to be accounted for. Lots of effort and a considerable number of methods have been devoted to study these effects from both the analytical and numerical bent. A selection of such methods, apart from the $1/N_c$ expansion \cite{djm95,dai,rfm00} and chiral perturbation theory (heavy, conventional, and relativistic) with different renormalization schemes \cite{jm255,jm259,jen91,b1,b2,lacour,sau1,sau2,sch,fuchs,geng1,geng2}, one also finds the chiral quark-soliton model \cite{suh}, chiral quark constituent model \cite{dahiya}, and the fast-growing lattice QCD \cite{pndme,rqcd,ale,pacs,dju,nme,gu}; relevant lattice data are reviewed by the Flavour Lattice Averaging Group \cite{flag}.

This paper is organized as follows. Section \ref{sec:axialc} presents some basic material about the axial vector operator in the $1/N_c$ expansion to set notation and conventions; a scrutinized derivation of that operator is discussed. Section \ref{sec:onel} provides a review about one-loop corrections in large-$N_c$ chiral perturbation theory; a novel approach to deal with baryon matrix elements of these corrections, exploiting the Clebsch-Gordan structures that come along loop graphs, is used to get all contributions present for $N_c=3$, which is useful to express results in terms of the coupling constants $D$, $F$, $\mathcal{C}$, and $\mathcal{H}$. Section \ref{sec:psb} discusses a revisited PSB analysis with the use of flavor projection operators. For this task, the most general operator basis constituted by linearly independent 3-body operators is constructed; flavor projection operators are applied to that basis to classify operators according to their transformation properties under a given $SU(3)$ representation. Section \ref{sec:akcop} provides analytical expressions for axial vector couplings for different channels. Section \ref{sec:num} presents numerical results obtained through a least-squares fit to data. Some implications are discussed. To close, Sec.~\ref{sec:con} presents some concluding remarks. The paper is complemented by two appendices. In Appendix \ref{app:sb} the full expressions containing contributions of explicit symmetry breaking to the axial couplings for the processes of interest are listed and their relation with low-energy constants of order $p^3$ found in the literature are discussed in Appendix \ref{app:lecs}.

\section{\label{sec:axialc}Baryon axial vector current in large-$N_c$ QCD}

In this section, some relevant facts on the large-$N_c$ limit of QCD are reviewed to set notation and conventions. The groundwork on the subject can be found in the original papers \cite{tHooft,ven,witten,dm1,djm94,djm95} and references therein.

The $1/N_c$ expansion of any baryon operator transforming according to a given $SU(2)\times SU(3)$ representation can be expressed as \cite{djm95}
\begin{equation}
\mathcal{O} = \sum_n c_n \frac{1}{N_c^{n-1}} \mathcal{O}_n, \label{eq:1ncE}
\end{equation}
where the $\mathcal{O}_n$ make up a complete set of linearly independent effective $n$-body operators which are written as polynomials in the $SU(6)$ generators of spin $J^i$, flavor $T^a$, and spin-flavor $G^{ia}$
\begin{subequations}
\label{eq:su6gen}
\begin{eqnarray}
& & J^i = q^\dagger \frac{\sigma^i}{2} q, \qquad \qquad \qquad \,\, (1,1) \\
& & T^a = q^\dagger \frac{\lambda^a}{2} q, \qquad \qquad \qquad (0,8) \\
& & G^{ia} = q^\dagger \frac{\sigma^i}{2}\frac{\lambda^a}{2} q, \,\,\,\quad \quad \qquad (1,8)
\end{eqnarray}
\end{subequations}
whose transformation properties under $SU(2)\times SU(3)$ are indicated as $(j,\mathrm{dim})$ in Eq.~(\ref{eq:su6gen}). The $SU(6)$ generators satisfy the commutation relations given in Table \ref{tab:surel} \cite{djm95}
\begingroup
\begin{table}
\caption{\label{tab:surel}$SU(6)$ commutation relations.}
\bigskip
\label{tab:su2fcomm}
\centerline{\vbox{ \tabskip=0pt \offinterlineskip
\halign{
\strut\quad $ # $\quad\hfil&\strut\quad $ # $\quad \hfil\cr
\multispan2\hfil $\left[J^i,T^a\right]=0,$ \hfil \cr
\noalign{\medskip}
\left[J^i,J^j\right] = i\epsilon^{ijk} J^k,
& \left[T^a,T^b\right] = i f^{abc} T^c,\cr
\noalign{\medskip}
\left[J^i,G^{ja}\right] = i\epsilon^{ijk} G^{ka},
&\left[T^a,G^{ib}\right] = i f^{abc} G^{ic},\cr
\noalign{\medskip}
\multispan2\hfil$\displaystyle [G^{ia},G^{jb}] = \frac{i}{4} \delta^{ij} f^{abc} T^c + \frac{i}{2N_f} \delta^{ab} \epsilon^{ijk} J^k + \frac{i}{2} \epsilon^{ijk} d^{abc} G^{kc}.$ \hfill\cr}}}
\end{table}
\endgroup
For definiteness, $q^\dagger$ and $q$ are $SU(6)$ operators that create and annihilate states in the fundamental representation of $SU(6)$, and $\sigma^k$ and $\lambda^c$ are the Pauli spin and Gell-Mann flavor matrices, respectively.

The baryon chiral Lagrangian describing the interactions of the pseudoscalar mesons and baryons in terms of QCD baryon operators has been discussed in detail in Ref.~\cite{jen96}. Each of these operators possesses an expansion in $1/N_c$ given by Eq.~(\ref{eq:1ncE}). Of particular interest in the present analysis is the flavor octet baryon axial vector current $A^{ia}$,
\begin{equation}
A^{ia} = \left\langle \mathcal{B}^\prime\left| \left(\overline{q} \gamma^i \gamma_5 \frac{\lambda^a}{2} q \right)_\mathrm{QCD} \right| \mathcal{B} \right\rangle.
\end{equation}

The baryon axial vector current at tree-level, hereafter denoted by $A_\mathrm{tree}^{ia}$, is a spin-1 operator with one flavor index so it transforms as $(1,8)$ under $SU(2)\times SU(3)$. The construction of the $1/N_c$ expansion of $A_\mathrm{tree}^{ia}$ was presented in Ref.~\cite{djm95}. An equivalent, more formal way to obtain this expansion can be achieved using the most general operator basis $R^{(ij)(abc)}$ introduced in the analysis of baryon-meson scattering \cite{banda2}. This operator basis is constituted by 170 linearly independent spin-2 operators with three flavor indices, retaining up to 3-body operators. $A_\mathrm{tree}^{ia}$ can be easily obtained from $R^{(ij)(abc)}$ by contracting the spin and flavor indices on $R^{(ij)(abc)}$ using the spin and flavor invariant tensors $\epsilon^{ijk}$ and $\delta^{ab}$. A straightforward analysis yields the complete set of linearly independent 3-body operators
\begin{subequations}
\label{eq:uno}
\begin{eqnarray}
W_1^{kc} & = & G^{kc}, \\
W_2^{kc} & = & \mathcal{D}_2^{kc}, \\
W_3^{kc} & = & i \epsilon^{ijk} \{J^i, G^{jc}\}, \\
W_4^{kc} & = & \mathcal{D}_3^{kc}, \\
W_5^{kc} & = & \mathcal{O}_3^{kc}.
\end{eqnarray}
\end{subequations}

The baryon axial vector operator $A_\mathrm{tree}^{kc}$ is a Hermitian operator and is odd under time reversal and so are the operators $W_j^{kc}$. The $1/N_c$ expansion of $A_\mathrm{tree}^{kc}$ can thus be written as
\begin{equation}
A_\mathrm{tree}^{kc} = \sum_{n=1}^5 w_n \frac{1}{N_c^{n-1}} W_n^{kc},
\end{equation}
where $w_n$ are unknown coefficients. The matrix elements $[A^{kc}]_{B_1B_2}$ of the spatial components of the axial vector current at zero recoil between baryon states $B_1$ and $B_2$, which fall into the lowest-lying irreducible representation of contracted-$SU(6)$ spin-flavor symmetry (the spin-1/2 octet and spin-3/2 decuplet baryons), can be written for $N_c=3$ in terms of the octet and decuplet pion coupling constants $D$, $F$, $\mathcal{C}$, and $\mathcal{H}$, each of which is of order $\mathcal{O}(N_c)$ \cite{jm255,jm259}. In this regard, operator $W_3^{kc} = i \epsilon^{ijk} \{J^i, G^{jc}\}$ is an unusual one; it can be rewritten as $-[J^2,G^{kc}]$, which has nonzero matrix elements only between baryon states of different spins, {\it i.e.,} it is nontrivial only in decuplet baryon $\to$ octet baryon transitions. A simple analysis shows that $w_3$ can be reabsorbed into $\mathcal{C}$ simply as $\mathcal{C} \to \mathcal{C}-w_3$, so the form of the $1/N_c$ expansion of $A_\mathrm{tree}^{kc}$ actually left at $N_c=3$ is \cite{djm95}
\begin{equation}
A_\mathrm{tree}^{kc} = a_1 G^{kc} + b_2 \frac{1}{N_c} \mathcal{D}_2^{kc} + b_3 \frac{1}{N_c^2} \mathcal{D}_3^{kc} + c_3 \frac{1}{N_c^2} \mathcal{O}_3^{kc}, \label{eq:akc}
\end{equation}
where $a_1$, $b_2$, $b_3$ and $c_3$ are operator coefficients not determined by the theory. They are related to the flavor octet baryon-pion couplings by
\begin{subequations}
\label{eq:foc}
\begin{eqnarray}
a_1 & = & \frac32 D + \frac32 F + \frac16 \mathcal{H}, \\
b_2 & = & - 4 D + 6 F, \\
b_3 & = & \frac32 D - \frac92 F - \frac12 \mathcal{H}, \\
c_3 & = & - 3 D - 3 F - 2 \mathcal{C} - \frac13 \mathcal{H}.
\end{eqnarray}
\end{subequations}
Hereafter, the axial vector coupling is defined as
\begin{equation}
g^{B_1B_2} = [A^{kc}]_{B_1B_2},
\end{equation}
where the spin index $k$ is understood to be conventionally set to 3 and the flavor index $c$ runs from 1 to 8. Transitions of interest here are those for which $c=1 \pm i2$ and $c=4 \pm i5$, which define the $\Delta S=0$ and $|\Delta S|=1$ transitions, respectively, where $S$ denotes the strangeness.

\section{\label{sec:onel}One-loop corrections in large-$N_c$ chiral perturbation theory}

Flavor $SU(3)$ symmetry can be broken in the axial vector current in two ways: Implicitly through the integrals occurring in the one-loop corrections and explicitly through PSB. In this section the first source of symmetry breaking is briefly reviewed.

\begin{figure}[ht]
\scalebox{0.9}{\includegraphics{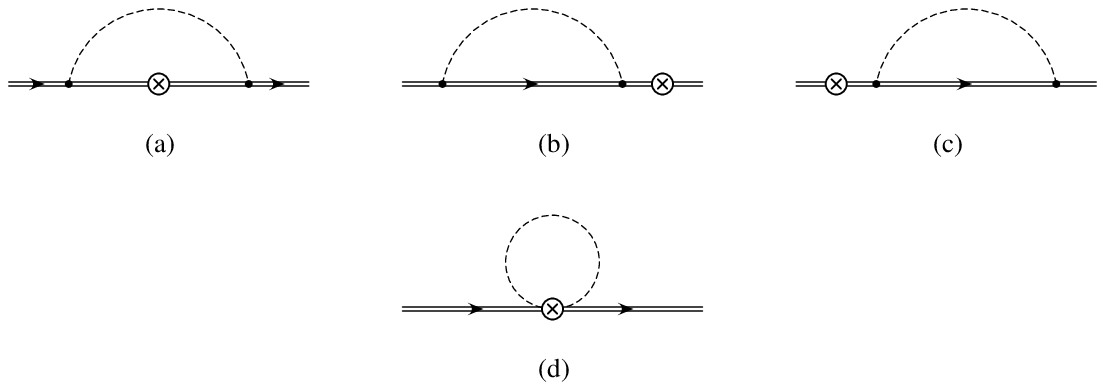}}
\caption{\label{fig:l1}One-loop corrections to the baryon axial vector current. A crossed circle represents an insertion of an axial vector current. Dashed lines and solid lines denote mesons and baryons, respectively.}
\end{figure}

The one-loop diagrams that renormalize the axial vector current are shown in Fig.~\ref{fig:l1}.

Diagrams \ref{fig:l1}(a,b,c) are noteworthy because they have a dependence on the ratio $\Delta/m$, where $\Delta$ denotes the decuplet-octet baryon mass difference and $m$ denotes the pseudoscalar meson mass. This dependence is manifest in a generic function $G(m,\Delta)$, which admits two power series expansions. On the one hand, in the chiral limit $m_q \to 0$ with $\Delta$ held fixed, $G(m,\Delta)$ can be expressed as \cite{rfm00}
\begin{equation}
G(m,\Delta) = G_0 + \left(\frac{m}{\Delta}\right) G_1 + \left(\frac{m}{\Delta}\right)^2 G_2 + \ldots
\end{equation}
On the other hand, in the limit $N_c \to \infty$ with $m$ held fixed, $G(m,\Delta)$ can be expressed as
\begin{equation}
G(m,\Delta) = \tilde{G}_0 + \left(\frac{\Delta}{m}\right) \tilde{G}_1 + \left(\frac{\Delta}{m}\right)^2 \tilde{G}_2 + \ldots
\end{equation}
The difference between these two expansions is usually referred to as the non-commutativity of the chiral and large-$N_c$ limits. Aspects of this subject were critically discussed for baryon isovector electric properties in Ref.~\cite{cohen}. However, the non-commutativity between the large-$N_c$ and chiral expansions does not necessarily mean that these limits are incompatible.
In order for chiral perturbation theory to be valid, the conditions $m \ll \Lambda_\chi$ and $\Delta \ll \Lambda_\chi$ should be fulfilled, where $\Lambda_\chi \sim 1$ GeV is the scale of chiral symmetry breaking. In the real world, $m_q \neq 0$ and $N_c \neq \infty$, so $G(m,\Delta)$ should be evaluated at the physical value of $\Delta/m$.

The contribution from Fig.~\ref{fig:l1}(a,b,c) to the renormalization of $A^{kc}$, after expanding the loop integral involved in a power series in $\Delta$, yields \cite{rfm00,rfm06,rfm12}
\begin{eqnarray}
\delta A_{1\mathrm{a}}^{kc} & = & \frac12 \left[A^{ja},\left[A^{jb},A^{kc}\right]\right] \Pi_{(1)}^{ab} - \frac12 \left\{ A^{ja}, \left[A^{kc},\left[\mathcal{M},A^{jb}\right] \right] \right\} \Pi_{(2)}^{ab} \nonumber \\
& & \mbox{} + \frac16 \left(\left[A^{ja}, \left[\left[\mathcal{M}, \left[ \mathcal{M},A^{jb}\right]\right],A^{kc}\right] \right] - \frac12 \left[\left[\mathcal{M},A^{ja}\right], \left[\left[\mathcal{M},A^{jb}\right],A^{kc}\right]\right]\right) \Pi_{(3)}^{ab} + \ldots \, , \label{eq:1a}
\end{eqnarray}
where the axial vector operators in Eq.~(\ref{eq:1a}) should be understood to be given at tree-level as expressed in Eq.~(\ref{eq:akc}), $\Pi_{(n)}^{ab}$ is a symmetric tensor which decomposes into flavor singlet, flavor $8$, and flavor $27$ representations, written as \cite{jen96}
\begin{eqnarray}
\Pi_{(n)}^{ab} = F_1^{(n)} \delta^{ab} + F_8^{(n)} d^{ab8} + F_{27}^{(n)} \left[ \delta^{a8} \delta^{b8} - \frac18 \delta^{ab} - \frac35 d^{ab8} d^{888}\right], \label{eq:pisym}
\end{eqnarray}
where
\begin{subequations}
\begin{eqnarray}
F_1^{(n)} & = & \frac18 \left[3F^{(n)}(m_\pi,0,\mu) + 4F^{(n)}(m_K,0,\mu) + F^{(n)}(m_\eta,0,\mu) \right], \label{eq:F1} \\
F_8^{(n)} & = & \frac{2\sqrt 3}{5} \left[\frac32 F^{(n)}(m_\pi,0,\mu) - F^{(n)}(m_K,0,\mu) - \frac12 F^{(n)}(m_\eta,0,\mu) \right], \label{eq:F8} \\
F_{27}^{(n)} & = & \frac13 F^{(n)}(m_\pi,0,\mu) - \frac43 F^{(n)}(m_K,0,\mu) + F^{(n)}(m_\eta,0,\mu). \label{eq:F27}
\end{eqnarray}
\end{subequations}

The functions $F_{\mathrm{dim}}^{(n)}$ depend on
\begin{equation}
F^{(n)}(m,\Delta,\mu) \equiv \frac{\partial^n F(m,\Delta,\mu)}{\partial \Delta^n},
\end{equation}
which is a basic loop integral with the full dependence on the ratio $\Delta/m$ and $\mu$ is the scale parameter of dimensional regularization. The loop integral is given by \cite{fmg}
\begin{eqnarray}
24\pi^2 f^2 F(m,\Delta,\mu) & = & -\Delta\left[\Delta^2-\frac32 m^2\right] \lambda_\epsilon + \Delta \left[\Delta^2-\frac32 m^2\right] \ln{\frac{m^2}{\mu^2}} - \frac83 \Delta^3 + \frac72 \Delta m^2 \nonumber \\
& & \mbox{} + \left\{ \begin{array}{ll}
\displaystyle 2(m^2-\Delta^2)^{3/2} \left[\frac{\pi}{2} - \tan^{-1} \left[\frac{\Delta}{\sqrt{m^2-\Delta ^2}}\right] \right], & |\Delta|< m \\[3mm]
\displaystyle - (\Delta^2-m^2)^{3/2} \ln \left[\frac{\Delta - \sqrt{\Delta^2-m^2}}{\Delta + \sqrt{\Delta^2-m^2}} \right], & |\Delta| > m. \end{array} \right. \label{eq:ib}
\end{eqnarray}
The $n$th derivatives of $F(m,\Delta,\mu)$ are readily obtained from Eq.~(\ref{eq:ib}). The expressions required here read
\begin{eqnarray}
16\pi^2f^2 F^{(1)}(m,\Delta,\mu) & = & (m^2-2\Delta^2)\left[ \lambda_\epsilon + 1 - \ln \frac{m^2}{\mu^2} \right] - 2\Delta^2 \nonumber \\
& & \mbox{} - \left\{ \begin{array}{ll} \displaystyle 4\Delta\sqrt{m^2-\Delta^2} \left[ \frac{\pi}{2}-\tan^{-1} \left[ \frac{\Delta}{\sqrt{m^2-\Delta^2}}\right]\right], & |\Delta|< m \\[6mm]
\displaystyle 2\Delta \sqrt{\Delta^2-m^2} \ln \left[ \frac{\Delta-\sqrt{\Delta^2-m^2}}{\Delta+\sqrt{\Delta^2-m^2}} \right], & |\Delta| > m. \end{array} \right. \label{eq:ibp}
\end{eqnarray}
\begin{equation}
4\pi^2f^2 F^{(2)}(m,\Delta,\mu) = -\Delta \left[ \lambda_\epsilon + 1 - \ln \frac{m^2}{\mu^2} \right] + \left\{ \begin{array}{ll} \displaystyle -\frac{m^2-2\Delta^2}{\sqrt{m^2-\Delta^2}} \left[ \frac{\pi}{2}-\tan^{-1} \left[ \frac{\Delta}{\sqrt{m^2-\Delta^2}}\right]\right], & |\Delta|< m \\[6mm] 
\displaystyle \frac{m^2-2\Delta^2}{2\sqrt{\Delta^2-m^2}} \ln \left[ \frac{\Delta-\sqrt{\Delta^2-m^2}}{\Delta+\sqrt{\Delta^2-m^2}}\right], & |\Delta| > m. \end{array} \right. \label{eq:ibpp}
\end{equation}
\begin{equation}
4\pi^2f^2 F^{(3)}(m,\Delta,\mu) = -\lambda_\epsilon -\frac{\Delta^2}{m^2-\Delta^2} + \ln \frac{m^2}{\mu^2} + \left\{ \begin{array}{ll} \displaystyle \frac{\Delta(3m^2-2\Delta^2)}{(m^2-\Delta^2)^{3/2}} \left[ \frac{\pi}{2} -\tan^{-1} \left[ \frac{\Delta}{\sqrt{m^2-\Delta^2}}\right] \right], & |\Delta|< m \\[6mm] 
\displaystyle \frac{\Delta(3m^2-2\Delta^2)}{2(\Delta^2-m^2)^{3/2}} \ln \left[ \frac{\Delta-\sqrt{\Delta^2-m^2}}{\Delta+\sqrt{\Delta^2-m^2}}\right], & |\Delta| > m.
\end{array} \right. \label{eq:ibppp}
\end{equation}

The renormalization of $A^{kc}$ in the combined formalism has been dealt with in Refs.~\cite{rfm06,rfm12,rfm21}. The aim of the analysis in Ref.~\cite{rfm06} was to provide the computational building blocks for more refined analyses including the decuplet-octet baryon mass difference effects studied in Ref.~\cite{rfm12} --limited to a few terms in the series-- to the evaluation of all contributions allowed at $N_c=3$ in the degeneracy limit $\Delta \to 0$ presented in Ref.~\cite{rfm21}. The latter specifically paved the way towards the present analysis.

From the computational point of view, the first summand in Eq.~(\ref{eq:1a}), $[A^{ja},[A^{jb},A^{kc}]]$, represents a major challenge. At $N_c=3$, the highest-order operator to be reduced is $N_c^{-6}[\mathcal{O}_3^{ja},[\mathcal{O}_3^{jb},\mathcal{O}_3^{kc}]]$, which is at most of order $\mathcal{O}(N_c)$, according to the large-$N_c$ cancellations discussed in Ref.~\cite{rfm00}. The pion decay constant $f\propto N_c$ so the function $F^{(1)}(m,0,\mu)$ is order $\mathcal{O}(N_c^{-1})$. Thus, the overall contribution of this term to $A^{kc}$ is $\mathcal{O}(N_c^0)$, or equivalently, $1/N_c$ times the tree-level value, which is order $\mathcal{O}(N_c)$.

Reference \cite{rfm21} presents the full evaluation of the operator structure $[A^{ja},[A^{jb},A^{kc}]]$ at $N_c=3$; the complete operator reductions are listed in Appendix A of that reference. The urge for the full computation sprang from the feasibility of carrying out a comparison with the HBChPT results for the axial vector coupling for baryon semileptonic decays of Refs.~\cite{jm255,jm259}. The comparison was analytically performed and it was proved that both approaches totally agree at $N_c=3$ in the degeneracy limit, using expressions (\ref{eq:foc}) to relate the unknowns of the approaches.\footnote{Actually, in Refs.~\cite{jm255,jm259} the limit $m_u=m_d=0$ was implemented and the Gell-Mann--Okubo formula was used to relate $m_\eta^2$ to $\frac43 m_K^2$.}

As for the contributions with one and two mass insertions, corresponding to the second and third summands of Eq.~(\ref{eq:1a}), respectively, they have been partially evaluated in previous works \cite{rfm12,rfm21} due to the inherent difficulty in reducing them analytically. For instance, the hardest terms to be reduced are $N_c^{-7}\{\mathcal{O}_3^{ja}, [\mathcal{O}_3^{kc},[J^2,\mathcal{O}_3^{jb}]]\}$ and $N_c^{-8} [[J^2,\mathcal{O}_3^{ja}],[[J^2,\mathcal{O}_3^{jb}],\mathcal{O}_3^{kc}]]$, which are both at most of order $\mathcal{O}(N_c)$. The analytical computations of these terms are extremely difficult, particularly in the $27$ representation of the latter, because it requires the construction of an operator basis containing up to 9-body operators, which can be constituted by a few hundred operators (this is a very crude estimate). Thus, a pragmatic approach should be implemented to achieve those reductions. The goal in this analysis is to find such an approach.

To outline how to proceed, it should be first pointed out that a simplification can be gained in the analysis of Ref.~\cite{rfm21}, {\it at the level of matrix element computations}. This alternative method will be implemented to tackle the two additional summands in Eq.~(\ref{eq:1a}). For this aim, let
\begin{equation}
\delta g^{B_1B_2} = \left[ \frac12 [A^{ja},[A^{jb},A^{kc}]] \Pi_{(1)}^{ab} \right]_{B_1B_2}, \label{eq:adeg}
\end{equation}
represent the matrix element of the one-loop contribution to the axial vector coupling $g$ in the degeneracy limit and $B_i$ denote baryons which fall into either octet or decuplet multiplets. $\delta g^{B_1B_2}$ in turn can be written as $\sum_\mathrm{dim} \delta g_{(\mathrm{dim})}^{B_1B_2}$ for $\mathrm{dim}=1$, $8$, and $27$ flavor representations. For definiteness, let $B_1=n$, $B_2=p$, fix $c=1+i2$ and pick the singlet piece of Eq.~(\ref{eq:adeg}) as a case study. Thus, $\delta g_{(1)}^{np}$ can generically be written, in the most general way, as
\begin{eqnarray}
\left. \delta g_{(1)}^{np} \right|_{\Delta\to 0} & = & \big[ x_1 a_1^3 + x_2 a_1^2b_2 + x_3 a_1b_2^2 + x_4 b_2^3 + x_5 a_1^2b_3 + x_6 a_1b_2b_3 + x_7 b_2^2b_3 + x_8 a_1b_3^2 + x_9 b_2b_3^2 + x_{10} b_3^3 \nonumber \\
& & \mbox{} + x_{11} a_1^2c_3 + x_{12} a_1b_2c_3 + x_{13} b_2^2c_3 + x_{14} a_1b_3c_3 + x_{15} b_2b_3c_3 + x_{16} b_3^2c_3 + x_{17} a_1c_3^2 + x_{18} b_2c_3^2 \nonumber \\
& & \mbox{}+ x_{19} b_3c_3^2 + x_{20} c_3^3 \big] F_1^{(1)}, \label{eq:gnp1}
\end{eqnarray}
where $x_{k}$ are the matrix elements of the operators involved in the double commutator of Eq.~(\ref{eq:adeg}). Their explicit forms can be read off the corresponding expressions listed in Appendix A of Ref.~\cite{rfm21} and are not needed in the example.

In HBChPT, the equivalent result is
\begin{equation}
\left. \delta g_{(1)}^{np} \right|_{\Delta\to 0} = \left[ \frac92 D^3 + \frac{25}{6} D^2F + \frac{15}{2} DF^2 + \frac{21}{2} F^3 + \frac{1}{18} \mathcal{C}^2D + \frac12 \mathcal{C}^2F + \frac{20}{27} \mathcal{C}^2\mathcal{H} \right] F_1^{(1)}. \label{eq:gnp2}
\end{equation}

Now, by substituting relations (\ref{eq:foc}) into (\ref{eq:gnp1}) and solving for $x_k$ using the homogeneous system of equations that can be constructed\footnote{Constructing the homogeneous system of equations is straightforward: For instance, the term $\mathcal{H}^3$ does no appear in Eq.~(\ref{eq:gnp2}), so the linear combination of $x_j$ that comes along with it vanishes. There are 13 such equations.} out of Eq.~(\ref{eq:gnp2}), it is straightforward to show that the system can be fully solved if only $x_1$, $x_2$, $x_3$, $x_4$, $x_5$, $x_6$, and $x_8$ are known. The procedure can be repeated for $8$ and $27$ representations to complete the analysis for octet-octet baryon axial vector couplings.

As for decuplet-decuplet and decuplet-octet baryon axial vector couplings, the nonvanishing terms fall into the sets $\{\mathcal{C}^2D,\mathcal{C}^2F,\mathcal{C}^2\mathcal{H},\mathcal{H}^3\}$ and $\{\mathcal{C}D^2,\mathcal{C}F^2,\mathcal{C}\mathcal{H}^2,\mathcal{C}DF,\mathcal{C}\mathcal{H}D,\mathcal{C}\mathcal{H}F,\mathcal{C}^3\}$, respectively, so the homogeneous systems of equations can be constructed accordingly. Thus, only a few terms in expansion (\ref{eq:adeg}) are required to fully evaluate $\delta g^{B_1B_2}$ in the degeneracy limit. This fact can be rigorously verified because all $x_k$ are available \cite{rfm21}.

The approach outlined above can also be applied to the second and third summands in Eq.~(\ref{eq:1a}). Explicitly, for the canonical example,
\begin{eqnarray}
\left. \delta g_{(1)}^{np} \right|_{\Delta} & = & \big[ y_1 a_1^3 + y_2 a_1^2b_2 + y_3 a_1b_2^2 + y_4 a_1^2b_3 + y_5 a_1b_2b_3 + y_6 a_1b_3^2 + y_7 a_1^2c_3 + y_8 a_1b_2c_3 + y_9 b_2^2c_3 \nonumber \\
& & \mbox{} + y_{10} a_1b_3c_3 + y_{11} a_1c_3^2 + y_{12} b_2b_3c_3 + y_{13} b_2c_3^2 + y_{14} b_3^2c_3 + y_{15} b_3c_3^2 + y_{16}c_3^3 \big] F_1^{(2)}. \label{eq:gnp3}
\end{eqnarray}
Assuming that only the terms that fall into the set $\{D^3,D^2F,DF^2,F^3,\mathcal{C}^2D,\mathcal{C}^2F,\mathcal{C}^2\mathcal{H}\}$ are nontrivial, the solution of the resultant homogeneous system of equations yields that only the $y_1$, $y_2$, $y_3$, $y_4$, and $y_6$ coefficients are necessary to fully evaluate $\delta g_{(1)}^{np}$ with one mass insertion. This is also valid for the other flavor representations, and, most importantly, also for two mass insertions. Furthermore, equivalent set of equations can be constructed for decuplet-decuplet and decuplet-octet baryon axial vector couplings, using the arguments stated above.

At this point, $\delta g^{B_1B_2}$ can be fully known with up to two mass insertions, according to Eq.~(\ref{eq:1a}). All the necessary operator reductions are listed in the supplementary material to this paper for completeness.

Finally, the correction from Fig.~\ref{fig:l1}(d) to $A^{kc}$ has been dealt with in Ref.~\cite{rfm06}. Succinctly, it can be given as
\begin{equation}
\delta A_{1\mathrm{d}}^{kc} = - \frac12 [T^a,[T^b,A^{kc}]]\Pi^{ab}, \label{eq:1d}
\end{equation}
where $\Pi^{ab}$ is a symmetric tensor whose structure is similar to Eq.~(\ref{eq:pisym}), except that the loop integral is now \cite{fmg}
\begin{equation}
I(m,\mu) = \frac{m^2}{16\pi^2f^2} \left[ - \lambda_\epsilon - 1 + \log\frac{m^2}{\mu^2} \right].
\end{equation}

To summarize, one-loop corrections to the baryon axial vector operator are written as
\begin{equation}
\delta A_L^{kc} = \delta A_{1\mathrm{a}}^{kc} + \delta A_{1\mathrm{d}}^{kc}, \label{eq:oneL}
\end{equation}
where $\delta A_{1\mathrm{a}}^{kc}$ and $\delta A_{1\mathrm{d}}^{kc}$ comprise the contributions from Fig.~\ref{fig:l1}(a,b,c) and Fig.~\ref{fig:l1}(d), respectively, and are given by Eqs.~(\ref{eq:1a}) and (\ref{eq:1d}). The matrix elements required to evaluate axial vector couplings for various processes can be found in Ref.~\cite{rfm21} and references therein.

To close this section, the fact that the present calculation exploits the near degeneracy between octet and decuplet baryons should be stressed. Notice that the loop integral $F^{(1)}(m,\Delta,\mu)$, Eq.~(\ref{eq:ibp}), admits two different limits, namely, $\Delta \ll m$ and $m \ll \Delta$. Expanding the integral for $\Delta \ll m$ and retaining only the leading terms yields,
\begin{subequations}
\begin{equation}
\frac{m^2}{16f^2\pi^2} \left[ \lambda_\epsilon \left(1-\frac{2\Delta^2}{m^2}\right) + \left( 1 -\log {\frac{m^2}{\mu^2}} \right) -2\pi \left(\frac{\Delta }{m}\right) + 2 \log{\frac{m^2}{\mu^2}} \left(\frac{\Delta}{m} \right)^2 \right]
\end{equation}
whereas the opposite limit $\Delta \gg m$ yields
\begin{equation}
\frac{m^2}{16f^2\pi^2} \left[ \lambda_\epsilon \left(1-\frac{2\Delta^2}{m^2} \right) + \left(1-\frac{2\Delta^2}{m^2} \right) \log \frac{\mu^2}{4\Delta^2} + \mathcal{O} \left[\frac{m}{\Delta}\right]^2 \log \frac{m^2}{4\Delta^2}+ \mbox{analytic terms in $m$} \right],
\end{equation}
\end{subequations}
From this result, it can be concluded that the terms $\log (\mu^2/4\Delta^2)$ and the analytical ones can be absorbed into local counterterms. The remaining infrared terms, which are non-analytical, are generated by higher dimension operators arising upon 
integrating out the decuplet. Therefore, in the chiral limit $\Delta \gg m$ so the decuplet cannot contribute to the nonanalytical corrections for octet processes since these corrections come from infrared divergences. The decuplet thus decouples in the large-$N_c$ limit. This decoupling was first pointed out in the analysis of baryon masses \cite{jen91} and later verified in the study of the baryon vector current \cite{fmg}.

\section{\label{sec:psb}Perturbative $SU(3)$ flavor symmetry breaking}

As it was discussed in Ref.~\cite{rfm21}, in the conventional chiral momentum counting scheme tree diagrams involving higher-order vertices and one-loop corrections contribute to the axial vector current alike. Some tree-diagram contributions are needed as counterterms for the divergent parts of the loop integrals, which introduce extra low-energy constants in the low-energy expansion. Due to the fact that the leading $SU(3)$ breaking effects of the axial vector current are order $\mathcal{O}(m_q)$, in the combined formalism terms of this order can be accounted for through PSB. In the present work, PSB is revisited in a more formal fashion by using the projection operator technique introduced in Ref.~\cite{banda1}.

The starting point of this analysis is to recall that flavor symmetry breaking in QCD is due to the light quark masses and transforms as a flavor octet. The baryon axial vector current $A^{kc}$ transforms as $(1,8)$ under $SU(2)\times SU(3)$. Thus, in order to account for PSB effects into $A^{kc}$, the operators contained in the decomposition
\begin{subequations}
\label{eq:8x8}
\begin{eqnarray}
(8 \otimes 8)_S & = & 1 \oplus 8 \oplus 27, \\
(8 \otimes 8)_A & = & 8 \oplus 10 \oplus \overline{10},
\end{eqnarray}
\end{subequations}
need be considered. This translates into looking for all linearly independent 3-body operators with spin 1 and two free flavor indices, one of which must be set to 8 to account for first-order $SU(3)$ symmetry breaking. A list of these operators was provided in Ref.~\cite{djm95}. Here, following an alternative approach, a more general operator basis can be obtained with the help of flavor projection operators \cite{banda1}. For this purpose, the most general basis of spin-1 operators with two flavor indices is constructed 
from the tensor $R^{(ij)(a_2e_1e_2)}$ introduced in Ref.~\cite{banda2}, contracting it with $\epsilon^{ijk} f^{a_1e_1e_2}$ or $i\epsilon^{ijk} d^{a_1e_1e_2}$. In either case, omitting numerical factors, the resultant basis is
\begin{equation}
{\sf S}^{ka_1a_2} = \{\tilde{S}_i^{ka_1a_2}\},
\end{equation}
where
\begin{eqnarray}
\label{eq:akab}
\begin{array}{lll}
\tilde{S}_{1}^{ka_1a_2} = \delta^{a_1a_2} J^k, & \quad \quad &
\tilde{S}_{2}^{ka_1a_2} = d^{a_1a_2e_1} G^{ke_1}, \\[2mm]
\tilde{S}_{3}^{ka_1a_2} = i f^{a_1a_2e_1} G^{ke_1}, & \quad \quad &
\tilde{S}_{4}^{ka_1a_2} = i \epsilon^{ijk} \{G^{ia_1},G^{ja_2}\}, \\[2mm]
\tilde{S}_{5}^{ka_1a_2} = \{G^{ka_1},T^{a_2}\}, & \quad \quad &
\tilde{S}_{6}^{ka_1a_2} = \{G^{ka_2},T^{a_1}\}, \\[2mm]
\tilde{S}_{7}^{ka_1a_2} = i \epsilon^{ijk} d^{a_1a_2e_1} \{J^i,G^{je_1}\}, & \quad \quad &
\tilde{S}_{8}^{ka_1a_2} = \epsilon^{ijk} f^{a_1a_2e_1} \{J^i,G^{je_1}\}, \\[2mm]
\tilde{S}_{9}^{ka_1a_2} = d^{a_1a_2e_1} \mathcal{D}_2^{ke_1}, & \quad \quad &
\tilde{S}_{10}^{ka_1a_2} = i f^{a_1a_2e_1} \mathcal{D}_2^{ke_1}, \\[2mm]
\tilde{S}_{11}^{ka_1a_2} = i f^{a_1a_2e_1} \mathcal{D}_3^{ke_1}, & \quad \quad &
\tilde{S}_{12}^{ka_1a_2} = d^{a_1a_2e_1} \mathcal{D}_3^{ke_1}, \\[2mm]
\tilde{S}_{13}^{ka_1a_2} = i f^{a_1a_2e_1} \mathcal{O}_3^{ke_1}, & \quad \quad &
\tilde{S}_{14}^{ka_1a_2} = d^{a_1a_2e_1} \mathcal{O}_3^{ke_1}, \\[2mm]
\tilde{S}_{15}^{ka_1a_2} = i \epsilon^{ijk} \{T^{a_1},\{J^i,G^{ja_2}\}\}, & \quad \quad &
\tilde{S}_{16}^{ka_1a_2} = i \epsilon^{ijk} \{T^{a_2},\{J^i,G^{ja_1}\}\}, \\[2mm]
\tilde{S}_{17}^{ka_1a_2} = \{J^k,\{T^{a_1},T^{a_2}\}\}, & \quad \quad &
\tilde{S}_{18}^{ka_1a_2} = \{J^k,\{G^{ra_1},G^{ra_2}\}\}, \\[2mm]
\tilde{S}_{19}^{ka_1a_2} = \delta^{a_1a_2} \{J^2,J^k\}, & \quad \quad &
\tilde{S}_{20}^{ka_1a_2} = \{G^{ka_1},\{J^r,G^{ra_2}\}\}, \\[2mm]
\tilde{S}_{21}^{ka_1a_2} = \{G^{ka_2},\{J^r,G^{ra_1}\}\}, & \quad \quad &
\end{array}
\end{eqnarray}
where $\tilde{S}_i^{ka_1a_2}$ constitute a complete set of linearly independent 3-body operators.

\subsection{Flavor projection operators}

Operators in the basis ${\sf S}^{ka_1a_2}$ possess flavor transformation properties according to decomposition (\ref{eq:8x8}). Each flavor representation can be projected out by means of flavor projection operators \cite{banda1,banda2}. This technique exploits the decomposition of the tensor space formed by the product of the adjoint space with itself $n$ times, $\prod_{i = 1}^n adj \otimes$, into subspaces labeled by a specific eigenvalue of the quadratic Casimir operator $C$ of $SU(3)$.

Thus, for the product of two $SU(3)$ adjoints, the flavor projectors $[\pr{\mathrm{dim}}]^{a_1a_2a_3a_4}$ for the irreducible representation of dimension $\mathrm{dim}$ contained in (\ref{eq:8x8}) read \cite{banda1},
\begin{equation}
[\pr{1}]^{a_1a_2a_3a_4} = \frac{1}{N_f^2-1} \delta^{a_1a_2} \delta^{a_3a_4}, \label{eq:p1ab}
\end{equation}
\begin{equation}
[\pr{8}]^{a_1a_2a_3a_4} = \frac{N_f}{N_f^2-4} d^{a_1a_2e_1} d^{a_3a_4e_1}, \label{eq:p8ab}
\end{equation}
\begin{equation}
[\pr{27}]^{a_1a_2a_3a_4} = \frac12 (\delta^{a_1a_3} \delta^{a_2a_4} + \delta^{a_2a_3} \delta^{a_1a_4}) - \frac{1}{N_f^2-1} \delta^{a_1a_2} \delta^{a_3a_4} - \frac{N_f}{N_f^2-4} d^{a_1a_2e_1} d^{a_3a_4e_1}, \label{eq:p27ab}
\end{equation}
\begin{equation}
[\pr{8_A}]^{a_1a_2a_3a_4} = \frac{1}{N_f} f^{a_1a_2e_1} f^{a_3a_4e_1}, \label{eq:p8aab}
\end{equation}
and
\begin{equation}
[\pr{10+\overline{10}}]^{a_1a_2a_3a_4} = \frac12 (\delta^{a_1a_3} \delta^{a_2a_4} - \delta^{a_2a_3} \delta^{a_1a_4}) - \frac{1}{N_f} f^{a_1a_2e_1} f^{a_3a_4e_1}, \label{eq:p10ab}
\end{equation}
which satisfy the completeness relation
\begin{equation}
[\mathcal{P}^{(1)} + \mathcal{P}^{(8)} + \mathcal{P}^{(27)} + \mathcal{P}^{(8_A)} + \mathcal{P}^{(10+\overline{10})}]^{a_1a_2a_3a_4} = \delta^{a_1a_3} \delta^{a_2a_4}.
\end{equation}

The action of flavor projection operators (\ref{eq:p1ab})-(\ref{eq:p10ab}) on the operators of the ${\sf S}^{ka_1a_2}$ basis can be better appreciated through some examples. Let $\tilde{S}_{21}^{ka_1a_2}$ be used to illustrate the procedure. Thus, the operators $[\mathcal{P}^{(\mathrm{dim})}\tilde{S}_{21}]^{ka_1a_2}$ read,
\begin{eqnarray}
[\mathcal{P}^{(1)}\tilde{S}_{21}]^{ka_1a_2} & = & \frac{1}{N_f^2-1} \delta^{a_1a_2} \{G^{ke_1},\{J^r,G^{re_1}\}\} \nonumber \\
& = & \frac{(N_c+2N_f-2)(N_c+2)}{4(N_f^2-1)} \delta^{a_1a_2} J^k + \frac{N_f-2}{2N_f(N_f^2-1)} \delta^{a_1a_2} \{J^2,J^k\},
\end{eqnarray}
\begin{eqnarray}
[\mathcal{P}^{(8)}\tilde{S}_{21}]^{ka_1a_2} & = & \frac{N_f}{N_f^2-4} d^{a_1a_2e_1} d^{e_2e_3e_1} \{G^{ke_3},\{J^r,G^{re_2}\}\} \nonumber \\
& = & \frac{N_f}{N_f+2} d^{a_1a_2e_1} G^{ke_1} + \frac{(N_c+N_f)N_f}{2(N_f^2-4)} d^{a_1a_2e_1} \mathcal{D}_2^{ke_1} + \frac{N_f-4}{2(N_f^2-4)} d^{a_1a_2e_1} \mathcal{D}_3^{ke_1} \nonumber \\
& & \mbox{} + \frac{1}{N_f+2} d^{a_1a_2e_1} \mathcal{O}_3^{ke_1},
\end{eqnarray}
\begin{eqnarray}
[\mathcal{P}^{(27)}\tilde{S}_{21}]^{ka_1a_2} & = & \frac12 \{G^{ka_1},\{J^r,G^{ra_2}\}\} + \frac12 \{G^{ka_2},\{J^r,G^{ra_1}\}\}
 - \frac{N_f}{N_f^2-4} d^{a_1a_2e_1} d^{e_2e_3e_1} \{G^{ke_3},\{J^r,G^{re_2}\}\} \nonumber \\
& & \mbox{} - \frac{1}{N_f^2-1} \delta^{a_1a_2} \{G^{ke_1},\{J^r,G^{re_1}\}\} \nonumber \\
& = & \frac12 \{G^{ka_1},\{J^r,G^{ra_2}\}\} + \frac12 \{G^{ka_2},\{J^r,G^{ra_1}\}\} - \frac{N_f}{N_f+2} d^{a_1a_2e_1} G^{ke_1} \nonumber \\
& & \mbox{} - \frac{(N_c+2N_f-2)(N_c+2)}{4(N_f^2-1)} \delta^{a_1a_2} J^k - \frac{(N_c+N_f)N_f}{2(N_f^2-4)} d^{a_1a_2e_1} \mathcal{D}_2^{ke_1} \nonumber \\
& & \mbox{} - \frac{N_f-4}{2(N_f^2-4)} d^{a_1a_2e_1} \mathcal{D}_3^{ke_1} - \frac{1}{N_f+2} d^{a_1a_2e_1} \mathcal{O}_3^{ke_1} - \frac{N_f-2}{2N_f(N_f^2-1)} \delta^{a_1a_2} \{J^2,J^k\},
\end{eqnarray}
\begin{eqnarray}
[\mathcal{P}^{(8_A)}\tilde{S}_{21}]^{ka_1a_2} & = & \frac{1}{N_f} f^{a_1a_2e_1} f^{e_2e_3e_1} \{G^{ke_3},\{J^r,G^{re_2}\}\} \nonumber \\
& = & \frac{N_c+N_f}{2N_f} \epsilon^{ijk} f^{a_1a_2e_1} \{J^i,G^{je_1}\},
\end{eqnarray}
and
\begin{eqnarray}
[\mathcal{P}^{(10+\overline{10})}\tilde{S}_{21}]^{ka_1a_2} & = & - \frac12 \{G^{ka_1},\{J^r,G^{ra_2}\}\} + \frac12 \{G^{ka_2},\{J^r,G^{ra_1}\}\} - f^{a_1a_2e_1} f^{e_2e_3e_1} \frac{1}{N_f} \{G^{ke_3},\{J^r,G^{re_2}\}\} \nonumber \\
& = & - \frac12 \{G^{ka_1},\{J^r,G^{ra_2}\}\} + \frac12 \{G^{ka_2},\{J^r,G^{ra_1}\}\} - \frac{N_c+N_f}{2 N_f} \epsilon^{ijk} f^{a_1a_2e_1} \{J^i,G^{je_1}\}.
\end{eqnarray}

Once the projection operators (\ref{eq:p1ab})-(\ref{eq:p10ab}) are applied to the operators of basis (\ref{eq:akab}), the resultant operators can be classified according to their transformation properties under $SU(3)$. These operators read \\

$1$ representation

\begin{equation}
S_{1,1}^{ka_1a_2} = \delta^{a_1a_2} J^k,
\end{equation}

\begin{equation}
S_{2,1}^{ka_1a_2} = \delta^{a_1a_2} \{J^2,J^k\},
\end{equation}

$8$ representation

\begin{equation}
S_{1,8}^{ka_1a_2} = d^{a_1a_2e_1} G^{ke_1},
\end{equation}

\begin{equation}
S_{2,8}^{ka_1a_2} = d^{a_1a_2e_1} \mathcal{D}_2^{ke_1},
\end{equation}

\begin{equation}
S_{3,8}^{ka_1a_2} = i d^{a_1a_2e_1} \epsilon^{ijk} \{J^i,G^{je_1}\},
\end{equation}

\begin{equation}
S_{4,8}^{ka_1a_2} = d^{a_1a_2e_1} \mathcal{D}_3^{ke_1},
\end{equation}

\begin{equation}
S_{5,8}^{ka_1a_2} = d^{a_1a_2e_1} \mathcal{O}_3^{ke_1},
\end{equation}

$27$ representation

\begin{equation}
S_{1,27}^{ka_1a_2} = \frac12 \{G^{ka_1},T^{a_2}\} + \frac12 \{G^{ka_2},T^{a_1}\} - \frac{N_c+N_f}{N_f(N_f+1)} \delta^{a_1a_2}J^k - \frac{N_c+N_f}{N_f+2} d^{a_1a_2e_1} G^{ke_1} - \frac{1}{N_f+2} d^{a_1a_2e_1} \mathcal{D}_2^{ke_1},
\end{equation}

\begin{eqnarray}
S_{2,27}^{ka_1a_2} & = & \{J^k,\{T^{a_1},T^{a_2}\}\} - \frac{N_c(N_c+2N_f)(N_f-2)}{N_f(N_f^2-1)} \delta^{a_1a_2}J^k - \frac{2(N_c+N_f)(N_f-4)}{N_f^2-4} d^{a_1a_2e_1} \mathcal{D}_2^{ke_1} \nonumber \\
& & \mbox{} - \frac{2N_f}{N_f^2-4} d^{a_1a_2e_1} \mathcal{D}_3^{ke_1} - \frac{2}{N_f^2-1} \delta^{a_1a_2} \{J^2,J^k\},
\end{eqnarray}

\begin{eqnarray}
S_{3,27}^{ka_1a_2} & = & \{J^k,\{G^{ra_1},G^{ra_2}\}\} - \frac{3N_c(N_c+2N_f)}{4(N_f^2-1)} \delta^{a_1a_2}J^k - \frac{3(N_c+N_f)N_f}{2(N_f^2-4)} d^{a_1a_2e_1} \mathcal{D}_2^{ke_1} \nonumber \\
& & \mbox{} + \frac{N_f+4}{2(N_f^2-4)} d^{a_1a_2e_1} \mathcal{D}_3^{ke_1} + \frac{N_f+2}{2N_f(N_f^2-1)} \delta^{a_1a_2} \{J^2,J^k\},
\end{eqnarray}

\begin{equation}
S_{4,27}^{ka_1a_2} = \frac12 i \epsilon^{ijk} \{T^{a_1},\{J^i,G^{ja_2}\}\} + \frac12 i \epsilon^{ijk} \{T^{a_2},\{J^i,G^{ja_1}\}\} - 
 \frac{N_c+N_f}{N_f+2} i \epsilon^{ijk} d^{a_1a_2e_1} \{J^i,G^{je_1}\},
\end{equation}

\begin{eqnarray}
S_{5,27}^{ka_1a_2} & = & \frac12 \{G^{ka_1},\{J^r,G^{ra_2}\}\} + \frac12 \{G^{ka_2},\{J^r,G^{ra_1}\}\} - \frac{(N_c+2)(N_c+2N_f-2)}{4(N_f^2-1)} \delta^{a_1a_2} J^k \nonumber \\
& & \mbox{} - \frac{N_f}{N_f+2} d^{a_1a_2e_1} G^{ke_1} - \frac{(N_c+N_f)N_f}{2(N_f^2-4)} d^{a_1a_2e_1} \mathcal{D}_2^{ke_1} - \frac{N_f-4}{2(N_f^2-4)} d^{a_1a_2e_1} \mathcal{D}_3^{ke_1} \nonumber \\
& & \mbox{} - \frac{1}{N_f+2} d^{a_1a_2e_1} \mathcal{O}_3^{ke_1} - \frac{N_f-2}{2N_f(N_f^2-1)} \delta^{a_1a_2} \{J^2,J^k\},
\end{eqnarray}

$8_A$ representation

\begin{equation}
S_{1,{8_A}}^{ka_1a_2} = i f^{a_1a_2e_1} G^{ke_1},
\end{equation}

\begin{equation}
S_{2,{8_A}}^{ka_1a_2} = i f^{a_1a_2e_1} \mathcal{D}_2^{ke_1},
\end{equation}

\begin{equation}
S_{3,{8_A}}^{ka_1a_2} = \epsilon^{ijk} f^{a_1a_2e_1} \{J^i,G^{je_1}\},
\end{equation}

\begin{equation}
S_{4,{8_A}}^{ka_1a_2} = i f^{a_1a_2e_1} \mathcal{D}_3^{ke_1},
\end{equation}

\begin{equation}
S_{5,{8_A}}^{ka_1a_2} = i f^{a_1a_2e_1} \mathcal{O}_3^{ke_1},
\end{equation}

$10 + \overline{10}$ representation

\begin{equation}
S_{1,10+\overline{10}}^{ka_1a_2} = \frac12 \{G^{ka_1},T^{a_2}\} - \frac12 \{G^{ka_2},T^{a_1}\} + \frac{1}{N_f} \epsilon^{ijk} f^{a_1a_2e_1} \{J^i,G^{je_1}\},
\end{equation}

\begin{equation}
S_{2,10+\overline{10}}^{ka_1a_2} = i \epsilon^{ijk}\{G^{ia_1},G^{ja_2}\} + \frac{N_c+N_f}{N_f} i f^{a_1a_2e_1} G^{ke_1} - \frac{1}{N_f} i f^{a_1a_2e_1} \mathcal{D}_2^{ke_1},
\end{equation}

\begin{equation}
S_{3,10+\overline{10}}^{ka_1a_2} = \frac12 \{G^{ka_1},\{J^r,G^{ra_2}\}\} - \frac12 \{G^{ka_2},\{J^r,G^{ra_1}\}\} + \frac{N_c+N_f}{2N_f} \epsilon^{ijk} f^{a_1a_2e_1} \{J^i,G^{je_1}\},
\end{equation}

\begin{equation}
S_{4,10+\overline{10}}^{ka_1a_2} = \frac12 i \epsilon^{ijk} \{T^{a_1},\{J^i,G^{ja_2}\}\} - \frac12 i \epsilon^{ijk} \{T^{a_2},\{J^i,G^{ja_1}\}\} + \frac{2}{N_f} i f^{a_1a_2e_1} \mathcal{O}_3^{ke_1}.
\end{equation}
Notice that $f^{a_1a_2e_2} S_{r,10+\overline{10}}^{ka_1a_2} = 0$.

As an important remark, it can be rigorously proved that operators $S_{r,\mathrm{dim}}^{ka_1a_2}$ make up a complete set of linearly independent Hermitian operators.

The $1/N_c$ expansions for PSB contributions to any baryon operator transforming as $(1,8)$ under $SU(2)\times SU(3)$ spin-flavor symmetry can be cast into,
\begin{equation}
\delta A_\mathrm{SB,\mathrm{1}}^{kc} = z_{1,\mathrm{1}} S_{1,\mathrm{1}}^{kc8} + z_{2,\mathrm{1}} \frac{1}{N_c^2} S_{2,\mathrm{1}}^{kc8}, \label{eq:sb1}
\end{equation}
\begin{equation}
\delta A_\mathrm{SB,\mathrm{8}}^{kc} = z_{1,\mathrm{8}} S_{1,\mathrm{8}}^{kc8} + \frac{1}{N_c} \sum_{j=2}^3 z_{j,\mathrm{8}} S_{j,\mathrm{8}}^{kc8} + \frac{1}{N_c^2} \sum_{j=4}^5 z_{j,\mathrm{8}} S_{j,\mathrm{8}}^{kc8},
\end{equation}
\begin{equation}
\delta A_\mathrm{SB,\mathrm{27}}^{kc} = z_{1,\mathrm{27}} \frac{1}{N_c} S_{1,\mathrm{27}}^{kc8} + \frac{1}{N_c^2} \sum_{j=2}^5 z_{j,\mathrm{27}} S_{j,\mathrm{27}}^{kc8},
\end{equation}
\begin{equation}
\delta A_\mathrm{SB,\mathrm{8_A}}^{kc} = z_{1,\mathrm{8_A}} S_{1,\mathrm{8_A}}^{kc8} + \frac{1}{N_c} \sum_{j=2}^3 z_{j,\mathrm{8_A}} S_{j,\mathrm{8_A}}^{kc8} + \frac{1}{N_c^2} \sum_{j=4}^5 z_{j,\mathrm{8_A}} S_{j,\mathrm{8_A}}^{kc8},
\end{equation}
\begin{equation}
\delta A_\mathrm{SB,\mathrm{10+\overline{10}}}^{kc} = \frac{1}{N_c} \sum_{j=1}^2 z_{j,\mathrm{10+\overline{10}}} S_{j,\mathrm{10+\overline{10}}}^{kc8} + \frac{1}{N_c^2 } \sum_{j=3}^4 z_{j,\mathrm{10+\overline{10}}} S_{j,\mathrm{10+\overline{10}}}^{kc8}, \label{eq:sb10}
\end{equation}
where terms up to order $\mathcal{O}(N_c^{-2})$ have been retained.

Thus, first-order SB contributions to $A^{kc}$ can be written as
\begin{equation}
\delta A_\mathrm{SB}^{kc} = \sum_\mathrm{dim} \delta A_\mathrm{SB,\mathrm{dim}}^{kc}. \label{eq:sba}
\end{equation}
for $\mathrm{dim}=1,8,8_A,27$, and $10+\overline{10}$ and $z_{j,\mathrm{dim}}$ are 21 unknown coefficients. The explicit expressions for the matrix elements of $\delta A_\mathrm{SB}^{kc}$ for some processes of interest, $[\delta A_\mathrm{SB}^{kc}]_{B_1B_2}$, are listed in Appendix \ref{app:sb} for the sake of completeness.

\section{\label{sec:akcop}Universality of the baryon axial vector current $A^{kc}$}

Gathering together all partial contributions, the baryon axial vector operator $A^{kc}$ can be constructed as
\begin{equation}
A^{kc} = A_\mathrm{tree}^{kc} + \delta A_\mathrm{L}^{kc} + \delta A_\mathrm{SB}^{kc}, \label{eq:akcfinal}
\end{equation}
where $A_\mathrm{tree}^{kc}$ is the tree-level contribution, Eq.~(\ref{eq:akc}), whereas $\delta A_\mathrm{L}^{kc}$ and $\delta A_\mathrm{SB}^{kc}$ arise from one-loop contributions, Eq.~(\ref{eq:oneL}), and PSB, Eq.~(\ref{eq:sba}), respectively. It should be remarked that Eq.~(\ref{eq:akcfinal}) will be given at $N_c=3$.

The matrix elements of the space components of the baryon axial vector current between initial and final baryon states $B_1$ and $B_2$ at zero recoil, $[A^{kc}]_{B_1B_2}$, with different choices of baryon states and suitable values of flavor $c$, yield appropriate baryon axial vector couplings; by convention, $k$ is fixed to 3. Hence the property of universality of the $A^{kc}$ operator becomes manifest. Some axial vector couplings are discussed in the following sections.

\subsection{$A^{kc}$ acting diagonally on spin-1/2 baryon states: $g_A^{B_1B_2}$ coupling}

When $B_1$ and $B_2$ fall into the spin-$1/2$ octet baryon sector, with $B_1\neq B_2$, the flavor $c$ can be fixed to $c=1 \pm i2$ or $c=4 \pm i5$, which defines the $\Delta S=0$ and $|\Delta S|=1$ processes. The axial vector coupling thus corresponds to $g_A^{B_1B_2}$ as defined in baryon semileptonic decays, normalized in such a way that $g_A^{np} \approx 1.27$ for neutron beta decay. Processes of interest are $n\to p$, $\Sigma^ \pm \to \Lambda$, $\Lambda \to p$, $\Sigma^- \to n$, $\Xi^-\to \Lambda$, $\Xi^-\to \Sigma^0$, and $\Xi^0\to \Sigma^+$, for which experimental information is available \cite{part}. The corresponding expressions are listed in Appendix \ref{app:sb}. Notice that if expressions (\ref{eq:ganp})--(\ref{eq:gaxzsp}) are worked out in the limits $m_u=m_d=0$, $m_\eta^2 \to \frac43 m_K^2$ and $\Delta \to 0$ and only the chiral logs in the loop integrals are retained, these expressions fully agree with their counterparts obtained within HBChPT presented in Refs.~\cite{jm255,jm259}.

\subsection{$A^{kc}$ not acting diagonally on baryon states: $g^{B_1B_2}$ coupling}

The off-diagonal axial vector couplings can be explored in the strong decay of a decuplet baryon into an octet baryon and a pion, namely, $\Delta \to N \pi$, $\Sigma^* \to \Lambda \pi$, $\Sigma^* \to \Sigma \pi$ and $\Xi^* \to \Xi \pi$. The axial vector couplings will be denoted by $g^{B_1B_2}$. Following the lines of Ref.~\cite{dai}, the couplings will be normalized in such a way that they become equal in the $SU(3)$ symmetric limit. The axial vector couplings for the above decays are also listed in Appendix \ref{app:sb}. Notice that if expressions (\ref{eq:gdn})--(\ref{eq:gxsx}) are worked out in the limits $m_u=m_d=0$, $m_\eta^2 \to \frac43 m_K^2$ and $\Delta \to 0$ and only the chiral logs in the loop integrals are retained, these expressions can be compared to their counterparts obtained within HBChPT presented in Ref.~\cite{butler2}. Additional global factors of $-\frac12$ and $\frac32$ in the Clebsch-Gordan coefficients that come along with the loop integrals $F^{(1)}(m_K)$ and $I(m_K)$, respectively, are found in the expressions of that reference with respect to the ones in the present work. This might suggest some missing symmetry factors in the corresponding Feynman diagrams in that reference.

\subsection{$A^{kc}$ acting diagonally on spin-3/2 baryon states: $g_A^{B_1B_2}$ coupling in Wu-type experiments}

The diagonal matrix elements of axial vector current $A^{kc}$ when $B_1$ and $B_2$ are both spin-3/2 baryon states can also be computed. In a recent publication \cite{wu}, Bertilsson and Leupold suggested a research program to conduct Wu-type experiments, {\it i.e.}, $B_1\to B_2 \ell^-\overline{\nu}_\ell$. Examples of energetically allowed processes are $\Delta^0 \to \Delta^+e^-\overline{\nu}_e$ and $\Omega^- \to {\Xi^*}^0\ell^-\overline{\nu}_\ell$, among others. The authors argued that while the former process is practically inaccessible mainly because $\Delta$ baryons are short-lived, the latter process, on the contrary, is feasible to be within the reach of experiments like BESIII or LHCb. They concluded that if the branching ratio can be determined with a precision of around $10\%$, it would be possible to estimate the size of the flavor baryon octet-pion coupling $\mathcal{H}$.

The semileptonic decays $\Delta^0 \to \Delta^+e^-\overline{\nu}_e$ and $\Omega^- \to {\Xi^*}^0\ell^-\overline{\nu}_\ell$ are clearly $\Delta S=0$ and $|\Delta S|=1$ processes, respectively. In the present formalism, the leading axial vector couplings $g_A^{\Delta^0\Delta^ +}$ and $g_A^{\Omega^-{\Xi^*}^0}$ are calculated and listed in Appendix \ref{app:sb}. A precise eventual determination of $g_A^{\Omega^-{\Xi^*}^0}$ will definitely impact on a precise determination of $\mathcal{H}$. In the meantime, it only can be extracted by existing data from a least-squares fit or via LQCD approaches.

\subsection{Proton matrix element of the eighth component of the axial current}

The proton matrix element of the eighth component of the axial current, denoted here as $g_8^{pp} = \left[A^{38}\right]_{pp}$, can also be evaluated in the present approach. The full expression is given in Appendix \ref{app:sb}. Expression (\ref{eq:a8}) can be compared to its counterpart computed in HBChPT \cite{jm255} under the limits $m_u=m_d=0$ and $m_\eta^2 \to \frac43 m_K^2$, keeping only chiral loops. Both expressions agree. In this regard, it is important to remark that $g_8^{pp}$ picks up a $z_{1,1}$ factor coming from the leading order term of the singlet PSB correction.

The matrix element $g_8^{pp}$ is quite important in hadronic physics to understand the structure of nucleon and nucleon excitations.
The European Muon Collaboration (EMC), prepared to analyze spin dependent muon-proton scattering experiments, found that, in addition to the up and down quarks, there are also other contributions to the proton spin \cite{ash1,ash2}.

\section{\label{sec:num}Numerical analysis}

A numerical analysis through a least-squares fit to the available experimental data can be performed to determine the unknown parameters in the approach. The unknowns are the operator coefficients $a_1$, $b_2$, $b_3$, and $c_3$ (or equivalently $D$, $F$, $\mathcal{C}$, and $\mathcal{H}$) from the axial vector current operator at tree-level Eq.~(\ref{eq:akc}), along with the 21 operator coefficients that accompany PSB introduced in Eq.~(\ref{eq:sba}). For the transitions dealt with here, there are further simplifications to help reduce the number of unknowns. Accordingly, the singlet operator coefficients can be left out; the coefficients $z_{3,8}$ and $z_{5,8}$ participate in baryon decuplet to baryon octet processes only and the latter can be absorbed into the former; this situation repeats itself with $z_{3,10+\overline{10}}$ and $z_{4,10+\overline{10}}$; similarly, $z_{3,27}$ can be absorbed into $z_{2,27}$. Finally, the terms from the $8_A$ representation are even under time reversal and although they are allowed for broken symmetry, with the available experimental data, no fit can be obtained with a physically acceptable solution when they are present so they will be suppressed. Thus, one is left with 15 free parameters.

The experimental information for octet baryons is given in terms of the decay rates $R$, the ratios $g_A/g_V$, the angular correlation coefficients $\alpha_{e\nu}$, and the spin-asymmetry coefficients $\alpha_e$, $\alpha_\nu$, $\alpha_B$, $A$, and $B$ \cite{part}. All eight decay rates and six possible $g_A/g_V$ ratios are measured; the ratios $g_A/g_V$ for $\Sigma^ \pm\to\Lambda$ semileptonic decays are undefined, so their $g_A$ couplings will be used instead, which can be obtained from their respective decay rates through a standard procedure. A summary of this experimental information can be found in Table II of Ref.~\cite{rfm12}, except that for neutron decay its decay rate should be updated to $(1.1384  \pm 0.0006) \times 10^{-3}\, \mathrm{s}^{-1}$ \cite{part}. For decuplet baryons, the axial couplings $g$ for the processes $\Delta \to N\pi$, $\Sigma^*\to\Lambda\pi$, $\Sigma^*\to\Sigma\pi$, and $\Xi^*\to\Xi\pi$ are given in Table IX of that reference too. Additional inputs are $f_\pi = 93\,\mathrm{MeV}$, $\mu = 1\, \mathrm{GeV}$, $\Delta = 0.237\, \mathrm{GeV}$ \cite{rfm24}, along with the pseudoscalar meson masses \cite{part}.

The theoretical expressions for the observables used in the analysis can be found in Ref.~\cite{rfm04}. The decay rates are written in terms of six form factors $f_i$ and $g_i$ ($i=1,\ldots,3$) which are functions of the momentum transfer squared $q^2$; for definiteness, $f_1(0) = g_V$, $g_1(0)=g_A$, $f_2$ is used at its $SU(3)$ symmetric value, $g_2=0$, and the contributions from $f_3$ and $g_3$ can be neglected for electron modes. The decay rates must also include both model-independent and model-dependent radiative corrections \cite{gk}.

The fitting procedure will be implemented by using $R$ and the angular correlation and spin-asymmetry coefficients along with the axial couplings $g$ to extract information on $a_1$, $b_2$, $b_3$, $c_3$, and $z_{k,\mathrm{dim}}$; using the angular coefficients instead of the $g_A/g_V$ ratios, not only avoids inconsistencies but provides a more sensitive test.\footnote{$\alpha_k$ and $g_A$ cannot be used simultaneously because they are not independent quantities.} There are thus 29 pieces of data to fit 15 parameters. The number of free parameters may be an inconvenient for the fitting procedure because it is quite likely to get unphysical solutions characterized by unphysical or unwanted values of certain parameters. Thus, some criteria are imposed on the solution to assess the success of a fit: {\it i)} The value of $\chi^2/\mathrm{dof}$ should be around 1, which requires the inclusion of theoretical uncertainties added in quadrature to the experimental errors; {\it ii)} the best-fit parameters should yield consistent values of the $SU(3)$ invariants $D$, $F$, $\mathcal{C}$, and $\mathcal{H}$; {\it iii)} a positive definite error matrix must be obtained. Notice that the latter is the most stringent one.

In order to contrast between different outputs, three scenarios can be implemented. As a preliminary exercise, the first scenario replicates the approach of Refs.~\cite{jm255,jm259}, which used the decay rates and the axial couplings for $\Delta=0$, without including PSB contributions. The inconvenience here is that no counterterms for the divergent parts of the loop integrals are present. The best-fit parameters are listed in the second column of Table \ref{t:bestf}, with a corresponding nominal theoretical error of $0.2$ added in quadrature to the experimental errors. A more realistic fit with a better physical interpretation can be achieved including PSB in $g_A$ for $\Delta=0$ and $\Delta\neq 0$, and also including second-order symmetry breaking effects in $g_V$ \cite{fmg}; these fits are labeled as Fit 2 and Fit 3, respectively. A theoretical error of $0.15$ is added to the decay rates and asymmetry coefficients whereas a theoretical error of $0.04$ is well suited to the axial couplings; this choice is reasonable because terms of order $1/N_c^3$ and higher are being omitted in the axial couplings. Preliminary fits under these working assumptions yielded a negligible value of $z_{4,8}$, so this parameter can be left out and the free parameters are reduce by one. Fit 2 and Fit 3 yield the best-fit parameters listed in the third and fourth columns of Table \ref{t:bestf}, respectively.

\begingroup
\begin{table}
\caption{\label{t:bestf}Best-fit parameters for Fits 1, 2, and 3.}
\begin{center}
\begin{tabular}{lrrr}
\hline\hline
                         &            Fit 1 &            Fit 2 &            Fit 3 \\
\hline
$a_1$                    & $ 0.98 \pm 0.06$ & $ 0.86 \pm 0.13$ & $ 0.92 \pm 0.19$ \\
$b_2$                    & $-0.11 \pm 0.09$ & $ 0.07 \pm 0.27$ & $-0.81 \pm 0.95$ \\
$b_3$                    & $ 0.57 \pm 0.26$ & $ 0.51 \pm 0.32$ & $ 0.73 \pm 0.49$ \\
$c_3$                    & $ 0.46 \pm 0.15$ & $-0.04 \pm 0.10$ & $-0.04 \pm 0.10$ \\
$z_{1,8}$                &                  & $-0.11 \pm 0.38$ & $-0.24 \pm 0.45$ \\
$z_{2,8}$                &                  & $ 1.72 \pm 1.70$ & $ 2.58 \pm 2.56$ \\
$z_{3,8}$                &                  & $-1.41 \pm 0.51$ & $-1.74 \pm 0.52$ \\
$z_{1,10+\overline{10}}$ &                  & $ 0.12 \pm 0.10$ & $ 0.13 \pm 0.10$ \\
$z_{2,10+\overline{10}}$ &                  & $ 0.13 \pm 0.10$ & $ 0.16 \pm 0.10$ \\
$z_{3,10+\overline{10}}$ &                  & $ 0.04 \pm 0.10$ & $ 0.04 \pm 0.10$ \\
$z_{1,27}$               &                  & $ 0.25 \pm 0.21$ & $ 0.06 \pm 0.20$ \\
$z_{2,27}$               &                  & $-0.10 \pm 0.10$ & $-0.09 \pm 0.10$ \\
$z_{4,27}$               &                  & $-0.10 \pm 0.10$ & $-0.09 \pm 0.10$ \\
$z_{5,27}$               &                  & $ 0.07 \pm 0.10$ & $ 0.06 \pm 0.10$ \\
\hline
$D$                      & $ 0.58 \pm 0.02$ & $ 0.51 \pm 0.03$ & $ 0.58 \pm 0.11$ \\
$F$                      & $ 0.37 \pm 0.02$ & $ 0.35 \pm 0.03$ & $ 0.25 \pm 0.13$ \\
$\mathcal{C}$            & $-1.21 \pm 0.05$ & $-0.84 \pm 0.13$ & $-0.90 \pm 0.19$ \\
$\mathcal{H}$            & $-2.72 \pm 0.60$ & $-2.66 \pm 0.64$ & $-1.98 \pm 0.39$ \\
$F/D$                    & $ 0.64 \pm 0.02$ & $ 0.69 \pm 0.08$ & $ 0.43 \pm 0.28$ \\
$3F-D$                   & $ 0.53 \pm 0.05$ & $ 0.55 \pm 0.10$ & $ 0.18 \pm 0.46$ \\
\hline
$\chi^2/\mathrm{dof}$    &           $ 0.8$ &           $1.7$  &            $1.6$ \\
\hline\hline
\end{tabular}
\end{center}
\end{table}
\endgroup

The entries listed in Table \ref{t:bestf} reveal some interesting findings. From the large-$N_c$ perspective, all operator coefficients $a_1$, $b_2$, $b_3$, $c_3$, and $z_{k,\mathrm{dim}}$ get values according to expectations, namely, they are roughly order $\mathcal{O}(N_c^0)$; however, large theoretical errors are obtained in some parameters, particularly in $z_{2,8}$, which is a resultant of the working assumptions. The criteria imposed on the fits are satisfied fairly well by all three fits. An exception can be found in criterion {\it (ii)}, which requires that the $SU(6)$ relations\footnote{The $SU(6)$ relations are easily obtained by retaining the leading order term of Eq.~(\ref{eq:akc})} $[F/D]_{SU(6)}=2/3$, $[\mathcal{C}]_{SU(6)}=-2D$, and $[\mathcal{H}]_{SU(6)}=-3D$ be consistently attained. This is not entirely fulfilled by $[\mathcal{C}]_{SU(6)}$ in Fits 2 and 3. As for $\mathcal{H}$, the $SU(6)$ relation is best suited in Fit 3, with a large theoretical uncertainty. The large shifts in $\mathcal{C}$ and $\mathcal{H}$ in Fits 2 and 3 with respect to Fit 1 are a reflection of one-loop corrections and the perceptible shifts in $D$ and $F$ between Fits 2 and 3 are a consequence of the inclusion of a non vanishing $\Delta$ in the latter.

The fact that $\mathcal{H}$ is determined entirely from loop corrections has important implications. To better appreciate this fact, the 
best-fit parameters from Fit 2 and 3 are used to evaluate the axial couplings for various processes. The results are listed in Table \ref{t:fit2} and \ref{t:fit3}. For each coupling, pieces corresponding to tree-level, perturbative and one-loop corrections are listed. For the latter, different corrections from flavor representations $1$, $8$, and $27$ are separated. There is a noticeably hierarchy among these flavor contributions so that the 27 representation is the most suppressed one, being the 1 representation the least suppressed one.

The poor determinations of $\mathcal{C}$ and $\mathcal{H}$ have an important impact on the tree-level contributions of the axial couplings. While the overall symmetry breaking corrections to $g_A^{B_1B_2}$ represent a fraction of the total value\footnote{In most hadronic quantities, $SU(3)$ breaking corrections are expected to amount to 20-30\% of the tree-level value.} so that the tree-level value is the most significant one, in $g^{B_1B_2}$ corrections due to symmetry breaking are as large as the tree-level ones in both Fit 2 and Fit 3, which may question the validity of the approach. Before drawing any conclusions, it should be kept in mind that the coupling ratios, but not necessarily their absolute normalization, are those predicted by $SU(6)$ symmetry. Accordingly, in Ref.~\cite{rfm00} it was concluded that \textit{the one-loop correction is very sensitive to the deviations of the axial vector coupling ratios from their $SU(6)$ values}. At tree-level, the $a_1$ term is the dominant contribution to $D$, $F$, $\mathcal{C}$, and $\mathcal{H}$ so the $c_3$ term is a $1/N_c^2$ correction. At one-loop level, however, $a_1^3$ and $c_3^3$ terms are just as important and should be treated on an equal footing.

\begingroup
\begin{table}
\caption{\label{t:fit2}Values of axial vector couplings $g^{B_1B_2}$ for vanishing $\Delta$ corresponding to the best-fit parameters from Fit 1. Specific $SU(3)$ flavor symmetry breaking contributions are displayed.}
\begin{center}
\begin{tabular}{lrrrrrrrrrrrrrrr}
\hline\hline
& & & & \multicolumn{3}{c}{Figs.~1(a)--(c),\,\,$\mathcal{O}(\Delta^0)$} & \multicolumn{3}{c}{Fig.~1(d)} \\
$B_1B_2$ & Total & Tree & SB & $1$ & $8$ & $27$ & $1$ & $8$ & $27$ \\ \hline
$np$                & $ 1.297$ & $ 0.867$ & $ 0.149$ & $ 0.127$ & $ 0.127$ & $ 0.000$ & $ 0.234$ & $-0.078$ & $ 0.002$ \\
$\Sigma^ \pm\Lambda$ & $ 0.564$ & $ 0.419$ & $-0.009$ & $ 0.073$ & $ 0.073$ & $ 0.000$ & $ 0.113$ & $-0.038$ & $ 0.001$ \\
$\Lambda p$         & $-0.917$ & $-0.643$ & $ 0.030$ & $-0.082$ & $-0.082$ & $-0.001$ & $-0.174$ & $-0.029$ & $ 0.004$ \\
$\Sigma^-n$         & $ 0.358$ & $ 0.160$ & $ 0.107$ & $ 0.052$ & $ 0.052$ & $ 0.000$ & $ 0.043$ & $ 0.007$ & $-0.001$ \\
$\Xi^-\Lambda$      & $ 0.234$ & $ 0.224$ & $-0.087$ & $ 0.010$ & $ 0.010$ & $-0.001$ & $ 0.060$ & $ 0.010$ & $-0.001$ \\
$\Xi^-\Sigma^0$     & $ 0.835$ & $ 0.613$ & $-0.062$ & $ 0.090$ & $ 0.090$ & $ 0.003$ & $ 0.166$ & $ 0.028$ & $-0.004$ \\
 $\Xi^0\Sigma^+$    & $ 1.181$ & $ 0.867$ & $-0.087$ & $ 0.127$ & $ 0.127$ & $ 0.004$ & $ 0.234$ & $ 0.039$ & $-0.005$ \\
$\Delta N$          & $-1.979$ & $-0.838$ & $-0.856$ & $-0.045$ & $-0.091$ & $ 0.005$ & $-0.226$ & $ 0.075$ & $-0.002$ \\
$\Sigma^*\Lambda$   & $-1.763$ & $-0.838$ & $-0.658$ & $-0.045$ & $-0.064$ & $-0.005$ & $-0.226$ & $ 0.075$ & $-0.002$ \\
$\Sigma^*\Sigma$    & $-1.580$ & $-0.838$ & $-0.736$ & $-0.045$ & $ 0.178$ & $ 0.013$ & $-0.226$ & $ 0.075$ & $-0.002$ \\
$\Xi^*\Xi$          & $-1.460$ & $-0.838$ & $-0.498$ & $-0.045$ & $ 0.084$ & $-0.010$ & $-0.226$ & $ 0.075$ & $-0.002$ \\
\hline
$\Omega^-\Xi^0$     & $-0.599$ & $-0.838$ & $ 0.528$ & $-0.045$ & $ 0.017$ & $-0.001$ & $-0.226$ & $-0.039$ & $ 0.005$ \\
$\Delta^0\Delta^+$  & $ 3.890$ & $ 1.773$ & $ 0.958$ & $ 1.138$ & $-0.293$ & $-0.009$ & $ 0.479$ & $-0.160$ & $ 0.004$ \\
$\Omega^-{\Xi^*}^0$ & $ 2.760$ & $ 1.536$ & $-0.472$ & $ 0.985$ & $ 0.248$ & $-0.012$ & $ 0.418$ & $ 0.069$ & $-0.009$ \\
$(pp)_8$            & $ 0.104$ & $ 0.158$ & $-0.127$ & $ 0.007$ & $ 0.007$ & $ 0.000$ & $ 0.043$ & $ 0.014$ & $ 0.003$ \\
\hline\hline
\end{tabular}
\end{center}
\end{table}
\endgroup

\begin{landscape}

\begingroup
\begin{table}
\caption{\label{t:fit3}Values of axial vector couplings $g^{B_1B_2}$ for nonvanishing $\Delta$ corresponding to the best-fit parameters from Fit 2. Specific $SU(3)$ flavor symmetry breaking contributions are displayed for each order in $\Delta$.}
\begin{center}
\small
\begin{tabular}{lrrrrrrrrrrrrrrr}
\hline\hline
& & & & \multicolumn{3}{c}{Figs.~1(a)--(c),\,\,$\mathcal{O}(\Delta^0)$} & \multicolumn{3}{c}{Figs.~1(a)--(c),\,\,$\mathcal{O}(\Delta)$} & \multicolumn{3}{c}{Figs.~1(a)--(c),\,\,$\mathcal{O}(\Delta^2)$} & \multicolumn{3}{c}{Fig.~1(d)} \\
$B_1B_2$ & Total & Tree & SB & $1$ & $8$ & $27$ & $1$ & $8$ & $27$ & $1$ & $8$ & $27$ & $1$ & $8$ & $27$ \\ \hline
$np$                  & $ 1.306$ & $ 0.832$ & $ 0.157$ & $ 0.113$ & $ 0.113$ & $ 0.002$ & $ 0.113$ & $-0.054$ & $-0.001$ & $ 0.005$ & $ 0.007$ & $ 0.000$ & $ 0.225$ & $-0.075$ & $ 0.002$ \\
$\Sigma^ \pm\Lambda$  & $ 0.520$ & $ 0.474$ & $-0.018$ & $ 0.125$ & $ 0.125$ & $-0.001$ & $-0.069$ & $-0.013$ & $ 0.000$ & $-0.060$ & $ 0.012$ & $ 0.000$ & $ 0.128$ & $-0.043$ & $ 0.001$ \\
$\Lambda p$           & $-0.909$ & $-0.546$ & $ 0.078$ & $-0.012$ & $-0.012$ & $ 0.001$ & $-0.207$ & $ 0.064$ & $-0.001$ & $-0.065$ & $-0.020$ & $ 0.001$ & $-0.147$ & $-0.025$ & $ 0.003$ \\
$\Sigma^-n$           & $ 0.347$ & $ 0.329$ & $ 0.165$ & $ 0.195$ & $ 0.195$ & $-0.001$ & $-0.281$ & $ 0.032$ & $ 0.000$ & $-0.151$ & $-0.022$ & $ 0.000$ & $ 0.089$ & $ 0.015$ & $-0.002$ \\
$\Xi^-\Lambda$        & $ 0.258$ & $ 0.072$ & $-0.100$ & $-0.113$ & $-0.113$ & $-0.002$ & $ 0.275$ & $-0.018$ & $ 0.002$ & $ 0.125$ & $ 0.014$ & $-0.002$ & $ 0.019$ & $ 0.003$ & $ 0.000$ \\
$\Xi^-\Sigma^0$       & $ 0.893$ & $ 0.589$ & $-0.066$ & $ 0.080$ & $ 0.080$ & $ 0.002$ & $ 0.080$ & $ 0.019$ & $-0.001$ & $ 0.003$ & $-0.002$ & $ 0.001$ & $ 0.159$ & $ 0.026$ & $-0.003$ \\
$\Xi^0\Sigma^+$       & $ 1.263$ & $ 0.832$ & $-0.093$ & $ 0.113$ & $ 0.113$ & $ 0.003$ & $ 0.113$ & $ 0.027$ & $-0.001$ & $ 0.005$ & $-0.003$ & $ 0.001$ & $ 0.225$ & $ 0.037$ & $-0.005$ \\
$\Delta N$            & $-1.982$ & $-0.896$ & $-0.966$ & $-0.111$ & $-0.041$ & $ 0.003$ & $ 0.140$ & $-0.069$ & $-0.001$ & $ 0.095$ & $ 0.029$ & $-0.001$ & $-0.242$ & $ 0.081$ & $-0.002$ \\
$\Sigma^*\Lambda$     & $-1.759$ & $-0.896$ & $-0.819$ & $-0.111$ & $-0.013$ & $-0.003$ & $ 0.140$ & $-0.008$ & $ 0.002$ & $ 0.095$ & $ 0.018$ & $-0.001$ & $-0.242$ & $ 0.081$ & $-0.002$ \\
$\Sigma^*\Sigma$      & $-1.599$ & $-0.896$ & $-0.751$ & $-0.111$ & $ 0.087$ & $ 0.009$ & $ 0.140$ & $-0.031$ & $-0.014$ & $ 0.095$ & $ 0.029$ & $ 0.008$ & $-0.242$ & $ 0.081$ & $-0.002$ \\
$\Xi^*\Xi$            & $-1.456$ & $-0.896$ & $-0.637$ & $-0.111$ & $ 0.065$ & $-0.009$ & $ 0.140$ & $ 0.042$ & $ 0.004$ & $ 0.095$ & $ 0.013$ & $ 0.002$ & $-0.242$ & $ 0.081$ & $-0.002$ \\
\hline
$\Omega^-\Xi^0$       & $-0.510$ & $-0.896$ & $ 0.557$ & $-0.111$ & $ 0.012$ & $-0.001$ & $ 0.140$ & $-0.041$ & $ 0.011$ & $ 0.095$ & $ 0.005$ & $-0.005$ & $-0.242$ & $-0.040$ & $ 0.005$ \\
$\Delta^0\Delta^+$    & $ 3.340$ & $ 1.323$ & $ 1.339$ & $ 0.468$ & $-0.104$ & $-0.002$ & $ 0.141$ & $ 0.016$ & $ 0.004$ & $-0.090$ & $ 0.001$ & $ 0.003$ & $ 0.357$ & $-0.119$ & $ 0.003$ \\ 
$\Omega^-{\Xi^*}^0$   & $ 1.485$ & $ 1.146$ & $-0.549$ & $ 0.406$ & $ 0.085$ & $-0.002$ & $ 0.128$ & $-0.010$ & $ 0.005$ & $-0.078$ & $ 0.005$ & $ 0.002$ & $ 0.309$ & $ 0.051$ & $-0.007$ \\ 
$(pp)_8$              & $ 0.063$ & $ 0.051$ & $-0.196$ & $-0.080$ & $ 0.032$ & $ 0.000$ & $ 0.195$ & $-0.098$ & $-0.001$ & $ 0.088$ & $ 0.053$ & $ 0.001$ & $ 0.014$ & $ 0.005$ & $ 0.001$ \\
\hline\hline
\end{tabular}
\end{center}
\end{table}
\endgroup

\end{landscape}

Some axial couplings can be provided from the best-fit parameters of Fit 2 and Fit 3, which correspond to two cases of interest, namely, $\Delta=0$ and $\Delta\neq 0$, respectively. Thus, in the first case,
\begin{subequations}
\label{eq:pd0}
\begin{eqnarray}
g_A^{\Omega^-\Xi^0} & = & -0.60 \pm 0.31, \\
g_A^{\Delta^0\Delta^+} & = & 3.29 \pm 1.41, \\
g_A^{\Omega^-{\Xi^*}^0} & = & 2.76 \pm 1.39, \\
g_8^{pp} & = & 0.10 \pm 0.07,
\end{eqnarray}
\end{subequations}
whereas in the second case,
\begin{subequations}
\label{eq:pd1}
\begin{eqnarray}
g_A^{\Omega^-\Xi^0} & = & -0.51 \pm 0.16, \\
g_A^{\Delta^0\Delta^+} & = & 3.34 \pm 1.56, \\
g_A^{\Omega^-{\Xi^*}^0} & = & 1.49 \pm 1.16, \\
g_8^{pp} & = & 0.06 \pm 0.14.
\end{eqnarray}
\end{subequations}
A word of caution is necessary here. The predicted axial couplings listed in Eqs.~(\ref{eq:pd0}) and (\ref{eq:pd1}) should be taken with care. The poor determinations of $\mathcal{C}$ and $\mathcal{H}$ have important effects on these determinations. Besides, most data on angular correlation and asymmetry coefficients are rather old, dating back from the 1980s. Future, improved measurements of those  coefficients and also an eventual experimental determination of $\mathcal{H}$ necessary will impact on th estimates listed in Eqs.~(\ref{eq:pd0}) and (\ref{eq:pd1}).

For completeness, the pattern of symmetry breaking for the vector coupling $g_V/g_V^{SU(3)}-1$ obtained using the best-fit parameters are listed in Table \ref{t:f1sb}, where a comparison with other formalisms is also presented \cite{fg,sha}. The agreement is acceptable. The pattern of symmetry breaking {\it decreases} the tree-level value of $g_V$ by a few percent.

\begingroup
\begin{table}
\caption{\label{t:f1sb}Symmetry breaking pattern $g_V/g_V^{SU(3)}-1$ for the vector coupling $g_V$.}
\begin{center}
\begin{tabular}{lrrrrr}
\hline\hline
                 &              Fit 2 &              Fit 3 &      Ref.~\cite{fg} &   LQCD \cite{sha} \\ \hline
$\Lambda p$      & $-0.045 \pm 0.003$ & $-0.050 \pm 0.005$ & $-0.067  \pm 0.015$ & $-0.05  \pm 0.02$ \\
$\Sigma^- n$     & $-0.029 \pm 0.006$ & $-0.019 \pm 0.003$ & $-0.025  \pm 0.010$ & $-0.02  \pm 0.03$ \\
$\Xi^- \Lambda$  & $-0.040 \pm 0.001$ & $-0.042 \pm 0.006$ & $-0.053  \pm 0.010$ & $-0.06  \pm 0.04$ \\
$\Xi^- \Sigma^0$ & $-0.046 \pm 0.004$ & $-0.043 \pm 0.010$ & $-0.068  \pm 0.017$ & $-0.05  \pm 0.02$ \\
$\Xi^0 \Sigma^+$ & $-0.046 \pm 0.004$ & $-0.043 \pm 0.010$ &                     &                   \\
\hline\hline
\end{tabular}
\end{center}
\end{table}
\endgroup

To close this section, it should be pointed out that a full lattice determination of all axial vector couplings in flavor $SU(3)$ symmetry is not available yet. Most analyses are focused to the axial-vector coupling of the nucleon, which has has proven to be notoriously challenging, mainly because the difficulties of overcoming the exponential noise problem and also of controlling the excited state contributions to the nucleon correlator. Different mechanisms have been implemented to deal with these difficulties. The averaged values presented for the isovector axial charge in the FLAG review \cite{flag} are $g_A^{u-d} = 1.263(10)$ and $g_A^{u-d} = 1.265(20)$ for $N_f=2+1+1$ and $N_f=2+1$, respectively. The $g_A$ for neutron beta decay listed in Tables \ref{t:fit2} and \ref{t:fit3} are fairly consistent to these lattice determination, mainly because the large theoretical errors used in those fits.

As a final remark, some of the explicit symmetry breaking terms included in the present analysis also play the role of counterterms for the divergent parts of the loop integrals. The analysis of the baryon axial coupling also performed in the combined formalism \cite{fg0}, where the so called $\xi$-expansion is implemented,\footnote{In the $\xi$-expansion, the $1/N_c$ and the low energy power countings are linked according to $1/N_c = \mathcal{O}(\xi) = \mathcal{O}(p)$.} presents a detailed determination of such counterterms. The identification of these terms with the ones presented here is not as direct as expected, mainly because the specific flavor representations ($1$, $8$, $10+\overline{10}$, $27$) introduced here make hard such a comparison, and also, not only leading but also subleading terms in $1/N_c$ are included here. The comparison requires an extra effort and will be attempted elsewhere.

A closer comparison can be better carried out with the expressions presented in Ref.~\cite{heo}, where the chiral Lagrangian with three-flavor baryon fields up to chiral order $p^3$ is presented, along with all low-energy constants that contribute to the axial-vector and pseudoscalar currents in the baryon octet and decuplet fields at next-to-leading order in the $1/N_c$ expansion. Remarks on this comparison are given in Appendix \ref{app:lecs}.

\section{\label{sec:con}Concluding remarks}

The objective of the present paper was to perform a calculation of the $SU(3)$ breaking corrections for the axial vector coupling constants for different channels to order $\mathcal{O}(p^2)$ in the chiral expansion and consistent with the constrains imposed by the $1/N_c$ expansion of QCD. The paper, so to speak, culminates a research program that started more than two decades ago to try to better understand the intricacies associated with the renormalization of baryon vector current in the combined formalism.

Early attempts set the groundwork and eventually got improvements with the inclusion of perturbative symmetry breaking effects along with contributions of baryon decuplet-octet mass difference. The present analysis provides the most complete expression for the baryon axial vector operator at $N_c=3$, which is universal in the sense that it can be applied to different channels depending on the matrix elements evaluated. To do so, the Clebsch-Gordan structures inherent in the various flavor contributions were related among them by solving homogeneous systems of equations exploiting the fact that only a few terms in the loop expansion are required to fully evaluate all contributions.

Immediate applications to estimate the leading axial vector couplings appearing in the semileptonic decays of decuplet baryons were performed. Specifically, for the semileptonic decays $\Omega^- \to \Xi^0 \ell^-\overline{\nu}_\ell$ and $\Omega^- \to {\Xi^*}^0\ell^-\overline{\nu}_\ell$. Predicted values are listed in Eqs.~(\ref{eq:pd0}) and (\ref{eq:pd1}) for $\Delta=0$ and $\Delta=0.237\, \mathrm{GeV}$, respectively. There are noticeably shifts from one determination and the other, so a conclusive value cannot be quoted at this moment. Future improved measurements of decay rates and asymmetry parameters, in particular in Wu-type experiments as described in Ref.~\cite{wu} may lead to a precise determination of the elusive coupling $\mathcal{H}$. This also will have an important impact on the analysis presented here. In the mean time, the values listed in Eqs.~(\ref{eq:pd0}) and (\ref{eq:pd1}) are only indicative.

\section*{Acknowledgement}
The authors are grateful to Consejo Nacional de Ciencia y Tecnolog{\'\i}a (Mexico) for support through the {\it Ciencia de Frontera} project CF-2023-I-162.

\appendix

\section{\label{app:sb}Explicit expressions for axial couplings}

The matrix elements of the axial vector operator $A^{kc}$ (\ref{eq:akcfinal}) between appropriate baryon states yield the corresponding axial couplings,
\begin{equation}
g^{B_1B_2} = g^{B_1B_2}_\mathrm{Tree} + \delta g^{B_1B_2}_\mathrm{L} + \delta g^{B_1B_2}_\mathrm{SB}. \label{eq:gAC}
\end{equation}

The different contributions are listed below.

\subsection{Tree-level contribution}

\begin{eqnarray}
g^{np}_{A,\mathrm{Tree}} & = & \frac56 a_1 + \frac16 b_2 + \frac{5}{18} b_3 \nonumber \\
& = & D + F,
\end{eqnarray}
\begin{eqnarray}
\sqrt{\frac{3}{2}} g^{\Sigma^ \pm \Lambda}_{A,\mathrm{Tree}} & = & \frac12 a_1 + \frac16 b_3 \nonumber \\
& = & D,
\end{eqnarray}
\begin{eqnarray}
\sqrt{6} g^{\Lambda p}_{A,\mathrm{Tree}} & = & - \frac32 a_1 - \frac12 b_2 - \frac12 b_3 \nonumber \\
& = & - D - 3 F
\end{eqnarray}
\begin{eqnarray}
g^{\Sigma^-n}_{A,\mathrm{Tree}} & = & \frac16 a_1 - \frac16 b_2 + \frac{1}{18} b_3 \nonumber \\
& = & D - F,
\end{eqnarray}
\begin{eqnarray}
\sqrt{6} g^{\Xi^-\Lambda}_{A,\mathrm{Tree}} & = & \frac12 a_1 + \frac12 b_2 + \frac16 b_3 \nonumber \\
& = & - D + 3 F,
\end{eqnarray}
\begin{eqnarray}
\sqrt{2} g^{\Xi^-\Sigma^0}_{A,\mathrm{Tree}} & = & \frac56 a_1 + \frac16 b_2 + \frac{5}{18} b_3 \nonumber \\
& = & D + F,
\end{eqnarray}
\begin{eqnarray}
g^{\Xi^0 \Sigma^+}_{A,\mathrm{Tree}} & = & \frac56 a_1 + \frac16 b_2 + \frac{5}{18} b_3 \nonumber \\
& = & D + F,
\end{eqnarray}
\begin{eqnarray}
g^{\Delta N}_\mathrm{Tree} & = & - a_1 - \frac12 c_3 \nonumber \\
& = & \mathcal{C},
\end{eqnarray}
\begin{eqnarray}
g^{\Sigma^*\Lambda}_\mathrm{Tree} & = & - a_1 - \frac12 c_3 \nonumber \\
& = & \mathcal{C},
\end{eqnarray}
\begin{eqnarray}
g^{\Sigma^*\Sigma}_\mathrm{Tree} & = & - a_1 - \frac12 c_3 \nonumber \\
& = & \mathcal{C},
\end{eqnarray}
\begin{eqnarray}
g^{\Xi^*\Xi}_\mathrm{Tree} & = & - a_1 - \frac12 c_3 \nonumber \\
& = & \mathcal{C},
\end{eqnarray}
\begin{eqnarray}
g^{\Omega^-\Xi^0}_\mathrm{A,Tree} & = & - a_1 - \frac12 c_3 \nonumber \\
& = & \mathcal{C},
\end{eqnarray}
\begin{eqnarray}
g^{\Delta^0 \Delta^+}_{A,\mathrm{Tree}} & = & a_1 + b_2 + \frac53 b_3 \nonumber \\
& = & - \frac23 \mathcal{H},
\end{eqnarray}
\begin{eqnarray}
\sqrt{3} g^{\Omega^-{\Xi^*}^0}_{A,\mathrm{Tree}} & = & \frac32 a_1 + \frac32 b_2 + \frac52 b_3 \nonumber \\
& = & - \mathcal{H},
\end{eqnarray}
\begin{eqnarray}
g^{pp\pi^0}_\mathrm{Tree} & = & \frac{5}{12} a_1 + \frac{1}{12} b_2 + \frac{5}{36} b_3 \nonumber \\
& = & \frac12 D + \frac12 F,
\end{eqnarray}
\begin{eqnarray}
\sqrt{3} g^{pp\eta}_\mathrm{Tree} & = & \frac14 a_1 + \frac14 b_2 + \frac{1}{12} b_3 \nonumber \\
& = & - \frac12 D + \frac32 F.
\end{eqnarray}

Notice that the above expressions have been written in terms of the operator coefficients $a_1$, $b_2$, $b_3$, and $c_3$ introduced in Ref.~(\ref{eq:akc}) and in terms of the $SU(3)$ invariants $D$, $F$, $\mathcal{C}$, and $\mathcal{H}$, which are related through Eq.~(\ref{eq:foc})

\subsection{One-loop contributions}

The one-loop contributions to the axial-couplings can be organized as\footnote{Hereafter, as a shorthand notation, $F^{(n)}(m)$ will stand for $F^{(n)}(m,0,\mu)$.}
\begin{eqnarray}
\delta g_A^{np} & = & \left[ 2 (D+F)^3 + \frac29 \mathcal{C}^2D + \frac29 \mathcal{C}^2F + \frac{50}{81} \mathcal{C}^2 \mathcal{H} \right] F^{(1)}(m_\pi) \nonumber \\
& & \mbox{} + \left[ \frac{13}{6} D^3 - \frac16 D^2F + \frac12 D F^2 + \frac{11}{2} F^3 - \frac16 \mathcal{C}^2
D + \frac{5}{18} \mathcal{C}^2F + \frac{10}{81} \mathcal{C}^2 \mathcal{H} \right] F^{(1)}(m_K) \nonumber \\
& & \mbox{} + \frac13 (D-3F)^2 (D+F) F^{(1)}(m_\eta) + \left[ \frac{10}{9} \mathcal{C}^2D + \frac{10}{9} \mathcal{C}^2F + \frac{50}{81} \mathcal{C}^2 \mathcal{H} \right] \Delta F^{(2)}(m_\pi) \nonumber \\
& & \mbox{} + \left[ \frac16 \mathcal{C}^2D + \frac{7}{18} \mathcal{C}^2F + \frac{10}{81} \mathcal{C}^2 \mathcal{H} \right] \Delta F^{(2)}(m_K) + \left[ \frac{19}{27} \mathcal{C}^2D + \frac{19}{27} \mathcal{C}^2F + \frac{25}{81} \mathcal{C}^2 \mathcal{H} \right] \Delta^2F^{(3)}(m_\pi) \nonumber \\
& & \mbox{} + \left[ \frac{5}{36} \mathcal{C}^2D + \frac{23}{108} \mathcal{C}^2F + \frac{5}{81} \mathcal{C}^2 \mathcal{H} \right] \Delta^2F^{(3)}(m_K) - (D+F) I(m_\pi) - \frac12 (D+F) I(m_K) + \ldots, \label{eq:ganp}
\end{eqnarray}

\begin{eqnarray}
\sqrt{6} \delta g_A^{\Sigma^ \pm \Lambda} & = & \left[ \frac{14}{3} D^3 + 2 D F^2 + \frac{29}{18} \mathcal{C}^2D - \frac43 \mathcal{C}^2F + \frac{10}{27} \mathcal{C}^2 \mathcal{H} \right] F^{(1)}(m_\pi) \nonumber \\
& & \mbox{} + \left[ 3 D^3 + 13 D F^2 + \frac89 \mathcal{C}^2D - \frac83 \mathcal{C}^2F + \frac{5}{27} \mathcal{C}^2 \mathcal{H} \right] F^{(1)}(m_K) + \left[ \frac43 D^3 - \frac16 \mathcal{C}^2D \right] F^{(1)}(m_\eta) \nonumber \\
& & \mbox{} + \left[ \frac{31}{18} \mathcal{C}^2D - \frac23 \mathcal{C}^2F + \frac{10}{27} \mathcal{C}^2 \mathcal{H} \right] \Delta F^{(2)}(m_\pi) + \left[ \frac{16}{9} \mathcal{C}^2D - \frac43 \mathcal{C}^2F + \frac{5}{27} \mathcal{C}^2 \mathcal{H} \right] \Delta F^{(2)}(m_K) \nonumber \\
& & \mbox{} + \frac16 \mathcal{C}^2D \Delta F^{(2)}(m_\eta) + \left[ \frac{95}{108} \mathcal{C}^2D - \frac29 \mathcal{C}^2F + \frac{5}{27} \mathcal{C}^2 \mathcal{H} \right] \Delta^2F^{(3)}(m_\pi) \nonumber \\
& & \mbox{} + \left[ \frac{28}{27} \mathcal{C}^2D - \frac49 \mathcal{C}^2F + \frac{5}{54} \mathcal{C}^2 \mathcal{H} \right] \Delta^2F^{(3)}(m_K) + \frac{5}{36} \mathcal{C}^2D \Delta^2F^{(3)}(m_\eta) - 2 D I(m_\pi) - D I(m_K) \nonumber \\
& & \mbox{} + \ldots,
\end{eqnarray}

\begin{eqnarray}
\sqrt{6} \delta g_A^{\Lambda p} & = & \left[ - \frac98 D^3 - \frac{81}{8} D^2F - \frac{75}{8} D F^2 - \frac{27}{8} F^3 + \frac{23}{12} \mathcal{C}^2D - \frac{17}{4} \mathcal{C}^2F - \frac{10}{9} \mathcal{C}^2 \mathcal{H} \right] F^{(1)}(m_\pi) \nonumber \\
& & \mbox{} + \left[ - \frac{31}{12} D^3 - \frac54 D^2F - \frac34 D F^2 - \frac{99}{4} F^3 + \frac14 \mathcal{C}^2
D - \frac54 \mathcal{C}^2F - \frac59 \mathcal{C}^2 \mathcal{H} \right] F^{(1)}(m_K) \nonumber \\
& & \mbox{} + \left[ - \frac{19}{24} D^3 - \frac98 D^2F + \frac{21}{8} D F^2 - \frac{27}{8} F^3 \right] F^{(1)}(m_\eta) + \left[ \frac{1}{12} \mathcal{C}^2D - \frac{19}{4} \mathcal{C}^2F - \frac{10}{9} \mathcal{C}^2 \mathcal{H} \right] \Delta F^{(2)}(m_\pi) \nonumber \\
& & \mbox{} + \left[ - \frac14 \mathcal{C}^2D - \frac74 \mathcal{C}^2F - \frac59 \mathcal{C}^2 \mathcal{H} \right] \Delta F^{(2)}(m_K) + \left[ - \frac{19}{72} \mathcal{C}^2D - \frac{59}{24} \mathcal{C}^2F - \frac59 \mathcal{C}^2 \mathcal{H} \right] \Delta^2F^{(3)}(m_\pi) \nonumber \\
& & \mbox{} + \left[ - \frac{5}{24} \mathcal{C}^2D - \frac{23}{24} \mathcal{C}^2F - \frac{5}{18} \mathcal{C}^2 \mathcal{H} \right] \Delta^2F^{(3)}(m_K) + \frac38 (D + 3F) I(m_\pi) + \frac34 (D+3F) I(m_K) \nonumber \\
& & \mbox{} + \frac38 (D+3F) I(m_\eta) + \ldots,
\end{eqnarray}

\begin{eqnarray}
\delta g_A^{\Sigma^-n} & = & \left[ \frac{35}{24} D^3 + \frac{23}{24} D^2F + \frac{11}{8} D F^2 - \frac{41}{8} F^3 + \frac{17}{18} \mathcal{C}^2D - \frac{41}{18} \mathcal{C}^2F - \frac{10}{81} \mathcal{C}^2 \mathcal{H} \right] F^{(1)}(m_\pi) \nonumber \\
& & \mbox{} + \left[ \frac{31}{12} D^3 - \frac{53}{12} D^2F + \frac{19}{4} D F^2 - \frac{17}{4} F^3 + \frac{35}{36} \mathcal{C}^2D - \frac{59}{36} \mathcal{C}^2F - \frac{5}{81} \mathcal{C}^2 \mathcal{H} \right] F^{(1)}(m_K) \nonumber \\
& & \mbox{} + \left[ \frac{11}{24} D^3 - \frac{17}{24} D^2F + \frac{11}{8} D F^2 - \frac98 F^3 + \frac{13}{36} \mathcal{C}^2D - \frac{7}{12} \mathcal{C}^2F \right] F^{(1)}(m_\eta) \nonumber \\
& & \mbox{} + \left[ \frac{19}{18} \mathcal{C}^2D - \frac{31}{18} \mathcal{C}^2F - \frac{10}{81} \mathcal{C}^2 \mathcal{H} \right] \Delta F^{(2)}(m_\pi) + \left[ \frac{37}{36} \mathcal{C}^2D - \frac{49}{36} \mathcal{C}^2F - \frac{5}{81} \mathcal{C}^2 \mathcal{H} \right] \Delta F^{(2)}(m_K) \nonumber \\
& & \mbox{} + \left[ \frac{11}{36} \mathcal{C}^2D - \frac{5}{12} \mathcal{C}^2F \right] \Delta F^{(2)}(m_\eta) + \left[ \frac{59}{108} \mathcal{C}^2D - \frac{83}{108} \mathcal{C}^2F - \frac{5}{81} \mathcal{C}^2 \mathcal{H} \right] \Delta^2F^{(3)}(m_\pi) \nonumber \\
& & \mbox{} + \left[ \frac{113}{216} \mathcal{C}^2D - \frac{137}{216} \mathcal{C}^2F - \frac{5}{162} \mathcal{C}^2 \mathcal{H} \right] \Delta^2F^{(3)}(m_K) + \left[ \frac{31}{216} \mathcal{C}^2D - \frac{13}{72} \mathcal{C}^2F \right] \Delta^2F^{(3)}(m_\eta) \nonumber \\
& & \mbox{} - \frac38 (D-F) I(m_\pi) - \frac34 (D - F) I(m_K) - \frac38 (D - F) I(m_\eta) + \ldots,
\end{eqnarray}

\begin{eqnarray}
\sqrt{6} \delta g_A^{\Xi^-\Lambda} & = & \left[ - \frac98 D^3 + \frac{81}{8} D^2F - \frac{75}{8} D F^2 + \frac{27}{8} F^3 - \frac43 \mathcal{C}^2D + 4 \mathcal{C}^2F + \frac59 \mathcal{C}^2 \mathcal{H} \right] F^{(1)}(m_\pi) \nonumber \\
& & \mbox{} + \left[ - \frac{31}{12} D^3 + \frac54 D^2F - \frac34 D F^2 + \frac{99}{4} F^3 - \frac94 \mathcal{C}^2D + \frac{19}{4} \mathcal{C}^2F + \frac59 \mathcal{C}^2 \mathcal{H} \right] F^{(1)}(m_K) \nonumber \\
& & \mbox{} + \left[ - \frac{19}{24} D^3 + \frac98 D^2F + \frac{21}{8} D F^2 + \frac{27}{8} F^3 - \frac{11}{12}
\mathcal{C}^2D + \frac34 \mathcal{C}^2F \right] F^{(1)}(m_\eta) \nonumber \\
& & \mbox{} + \left[ - \frac76 \mathcal{C}^2D + \frac72 \mathcal{C}^2F + \frac59 \mathcal{C}^2 \mathcal{H} \right] \Delta F^{(2)}(m_\pi) + \left[ - \frac74 \mathcal{C}^2D + \frac{17}{4} \mathcal{C}^2F + \frac59 \mathcal{C}^2 \mathcal{H} \right] \Delta F^{(2)}(m_K) \nonumber \\
& & \mbox{} + \left[ - \frac{7}{12} \mathcal{C}^2D + \frac34 \mathcal{C}^2F \right] \Delta F^{(2)}(m_\eta) + \left[ - \frac59 \mathcal{C}^2D + \frac53 \mathcal{C}^2F + \frac{5}{18} \mathcal{C}^2 \mathcal{H} \right] \Delta^2F^{(3)}(m_\pi) \nonumber \\
& & \mbox{} + \left[ - \frac{19}{24} \mathcal{C}^2D + \frac{49}{24} \mathcal{C}^2F + \frac{5}{18} \mathcal{C}^2 \mathcal{H} \right] \Delta^2F^{(3)}(m_K) + \left[ - \frac{17}{72} \mathcal{C}^2D + \frac38 \mathcal{C}^2F \right] \Delta^2F^{(3)}(m_\eta) \nonumber \\
& & \mbox{} - \frac38 (-D+3F) I(m_\pi) - \frac34 (-D+3F) I(m_K) - \frac38 (- D + 3F) I(m_\eta) + \ldots,
\end{eqnarray}

\begin{eqnarray}
\sqrt{2} \delta g_A^{\Xi^- \Sigma^0} & = & g_A^{\Xi^0 \Sigma^+} \nonumber \\
& = & \left[ \frac{35}{24} D^3 - \frac{23}{24} D^2F + \frac{11}{8} D F^2 + \frac{41}{8} F^3 - \frac{1}{36} \mathcal{C}^2D + \frac{7}{36} \mathcal{C}^2F + \frac{10}{81} \mathcal{C}^2 \mathcal{H} \right] F^{(1)}(m_\pi) \nonumber \\
& & \mbox{} + \left[ \frac{31}{12} D^3 + \frac{53}{12} D^2F + \frac{19}{4} D F^2 + \frac{17}{4} F^3 - \frac{1}{12} \mathcal{C}^2
D + \frac{5}{36} \mathcal{C}^2F + \frac{35}{81} \mathcal{C}^2 \mathcal{H} \right] F^{(1)}(m_K) \nonumber \\
& & \mbox{} + \left[ \frac{11}{24} D^3 + \frac{17}{24} D^2F + \frac{11}{8} D F^2 + \frac98 F^3 + \frac16 \mathcal{C}^2
D + \frac16 \mathcal{C}^2F + \frac{5}{27} \mathcal{C}^2 \mathcal{H} \right] F^{(1)}(m_\eta) \nonumber \\
& & \mbox{} + \left[ \frac{7}{36} \mathcal{C}^2D + \frac{11}{36} \mathcal{C}^2F + \frac{10}{81} \mathcal{C}^2 \mathcal{H} \right] \Delta F^{(2)}(m_\pi) + \left[ \frac34 \mathcal{C}^2D + \frac{31}{36} \mathcal{C}^2F + \frac{35}{81} \mathcal{C}^2 \mathcal{H} \right] \Delta F^{(2)}(m_K) \nonumber \\
& & \mbox{} + \left[ \frac13 \mathcal{C}^2D + \frac13 \mathcal{C}^2F + \frac{5}{27} \mathcal{C}^2 \mathcal{H} \right] \Delta F^{(2)}(m_\eta) + \left[ \frac{29}{216} \mathcal{C}^2D + \frac{37}{216} \mathcal{C}^2F + \frac{5}{81} \mathcal{C}^2 \mathcal{H} \right] \Delta^2F^{(3)}(m_\pi) \nonumber \\
& & \mbox{} + \left[ \frac{37}{72} \mathcal{C}^2D + \frac{119}{216} \mathcal{C}^2F + \frac{35}{162} \mathcal{C}^2 \mathcal{H} \right] \Delta^2F^{(3)}(m_K) + \left[ \frac{7}{36} \mathcal{C}^2D + \frac{7}{36} \mathcal{C}^2F + \frac{5}{54} \mathcal{C}^2 \mathcal{H} \right] \Delta^2F^{(3)}(m_\eta) \nonumber \\
& & \mbox{} - \frac38 (D+F) I(m_\pi) - \frac34 (D+F) I(m_K) - \frac38 (D+F) I(m_\eta) + \ldots, \label{eq:gaxzsp}
\end{eqnarray}

\begin{eqnarray}
\delta g^{\Delta N} & = & \left[ \frac{43}{36} \mathcal{C}^3 + \frac18 \mathcal{C} D^2 + \frac18 \mathcal{C} F^2 + \frac14 \mathcal{C} DF + \frac{25}{36} \mathcal{C} D \mathcal{H} + \frac{25}{36} \mathcal{C} F \mathcal{H} + \frac{25}{648} \mathcal{C} \mathcal{H}^2 \right] F^{(1)}(m_\pi) \nonumber \\
& & \mbox{} + \left[ \frac{17}{36} \mathcal{C}^3 + \frac{11}{12} \mathcal{C} D^2 + \frac{13}{4} \mathcal{C} F^2 - \frac72 \mathcal{C} DF + \frac59 \mathcal{C} F \mathcal{H} + \frac{25}{324} \mathcal{C} \mathcal{H}^2 \right] F^{(1)}(m_K) \nonumber \\
& & \mbox{} + \left[ \frac18 \mathcal{C} D^2 + \frac98 \mathcal{C} F^2 - \frac34 \mathcal{C} DF - \frac{5}{36} \mathcal{C} D
\mathcal{H} + \frac{5}{12} \mathcal{C} F \mathcal{H} + \frac{5}{72} \mathcal{C} \mathcal{H}^2 \right] F^{(1)}(m_\eta) \nonumber \\
& & \mbox{} + \left[ \frac34 \mathcal{C}^3 + \frac12 \mathcal{C} D^2 + \frac12 \mathcal{C} F^2 + \mathcal{C} DF - \frac{25}{162} \mathcal{C} \mathcal{H}^2 \right] \Delta F^{(2)}(m_\pi) \nonumber \\
& & \mbox{} + \left[ \frac16 \mathcal{C} D^2 - \frac12 \mathcal{C} F^2 + \mathcal{C} DF - \frac{5}{162} \mathcal{C} \mathcal{H}^2 \right]
\Delta F^{(2)}(m_K) \nonumber \\
& & \mbox{} + \left[ \frac{133}{216} \mathcal{C}^3 - \frac16 \mathcal{C} D^2 - \frac16 \mathcal{C} F^2 - \frac13 \mathcal{C} D
F - \frac{25}{486} \mathcal{C} \mathcal{H}^2 \right] \Delta^2F^{(3)}(m_\pi) \nonumber \\
& & \mbox{} + \left[ \frac{53}{216} \mathcal{C}^3 - \frac{1}{18} \mathcal{C} D^2 + \frac16 \mathcal{C} F^2 - \frac13 \mathcal{C} DF - \frac{5}{486} \mathcal{C} \mathcal{H}^2 \right] \Delta^2F^{(3)}(m_K) - \mathcal{C} I(m_\pi) - \frac12 \mathcal{C} I(m_K) + \ldots, \label{eq:gdn}
\end{eqnarray}

\begin{eqnarray}
\delta g^{\Sigma^*\Lambda} & = & \left[ \frac{67}{72} \mathcal{C}^3 + \frac56 \mathcal{C} D^2 - \frac43 \mathcal{C} DF + \frac{10}{27} \mathcal{C} D \mathcal{H} - \frac{5}{81} \mathcal{C} \mathcal{H}^2 \right] F^{(1)}(m_\pi) \nonumber \\
& & \mbox{} + \left[ \frac{11}{18} \mathcal{C}^3 + \frac12 \mathcal{C} D^2 + \frac92 \mathcal{C} F^2 - \frac{8}{3} \mathcal{C} DF + \frac{5}{27} \mathcal{C} D \mathcal{H} + \frac53 \mathcal{C} F \mathcal{H} + \frac{20}{81} \mathcal{C} \mathcal{H}^2 \right] F^{(1)}(m_K) \nonumber \\
& & \mbox{} + \left[ \frac18 \mathcal{C}^3 - \frac16 \mathcal{C} D^2 \right] F^{(1)}(m_\eta) + \left[ \frac{13}{24} \mathcal{C}^3 + \frac13 \mathcal{C} D^2 + \frac23 \mathcal{C} DF - \frac{10}{81} \mathcal{C} \mathcal{H}^2 \right] \Delta F^{(2)}(m_\pi) \nonumber \\
& & \mbox{} + \left[ \frac13 \mathcal{C}^3 + \frac43 \mathcal{C} DF - \frac{5}{81} \mathcal{C} \mathcal{H}^2 \right] \Delta F^{(2)}(m_K) + \left[ - \frac18 \mathcal{C}^3 + \frac13 \mathcal{C} D^2 \right] \Delta F^{(2)}(m_\eta) \nonumber \\
& & \mbox{} + \left[ \frac{205}{432} \mathcal{C}^3 - \frac19 \mathcal{C} D^2 - \frac29 \mathcal{C} DF - \frac{10}{243} \mathcal{C} \mathcal{H}^2 \right] \Delta^2F^{(3)}(m_\pi) \nonumber \\
& & \mbox{} + \left[ \frac{35}{108} \mathcal{C}^3 - \frac49 \mathcal{C} DF - \frac{5}{243} \mathcal{C} \mathcal{H}^2 \right] \Delta^2F^{(3)}(m_K) + \left[ \frac{1}{16} \mathcal{C}^3 - \frac19 \mathcal{C} D^2 \right] \Delta^2F^{(3)}(m_\eta) \nonumber \\
& & \mbox{} - \mathcal{C} I(m_\pi) - \frac12 \mathcal{C} I(m_K) + \ldots,
\end{eqnarray}

\begin{eqnarray}
\delta g^{\Sigma^*\Sigma} & = & \left[ \frac{31}{72} \mathcal{C}^3 + \frac76 \mathcal{C} D^2 + \mathcal{C} F^2 - 2 \mathcal{C} DF + \frac59 \mathcal{C} D \mathcal{H} + \frac59 \mathcal{C} F \mathcal{H} + \frac{25}{81} \mathcal{C} \mathcal{H}^2 \right] F^{(1)}(m_\pi) \nonumber \\
& & \mbox{} + \left[ \frac{17}{18} \mathcal{C}^3 - \frac12 \mathcal{C} D^2 + \frac72 \mathcal{C} F^2 + \frac{10}{9} \mathcal{C} F \mathcal{H} - \frac{10}{81} \mathcal{C} \mathcal{H}^2 \right] F^{(1)}(m_K) \nonumber \\
& & \mbox{} + \left[ \frac{7}{24} \mathcal{C}^3 + \frac12 \mathcal{C} D^2 - 2 \mathcal{C} DF \right] F^{(1)}(m_\eta) + \left[ - \frac{1}{24} \mathcal{C}^3 - \frac13 \mathcal{C} D^2 + \mathcal{C} F^2 + \mathcal{C} DF + \frac{5}{81} \mathcal{C} \mathcal{H}^2 \right] \Delta F^{(2)}(m_\pi) \nonumber \\
& & \mbox{} + \left[ \frac23 \mathcal{C}^3 + \mathcal{C} D^2 - \mathcal{C} F^2 - \frac{20}{81} \mathcal{C} \mathcal{H}^2 \right] \Delta F^{(2)}(m_K) + \left[ \frac18 \mathcal{C}^3 + \mathcal{C} DF \right] \Delta F^{(2)}(m_\eta) \nonumber \\
& & \mbox{} + \left[ \frac{85}{432} \mathcal{C}^3 + \frac19 \mathcal{C} D^2 - \frac13 \mathcal{C} F^2 - \frac13 \mathcal{C} DF + \frac{5}{243} \mathcal{C} \mathcal{H}^2 \right] \Delta^2F^{(3)}(m_\pi) \nonumber \\
& & \mbox{} + \left[ \frac{53}{108} \mathcal{C}^3 - \frac13 \mathcal{C} D^2 + \frac13 \mathcal{C} F^2 - \frac{20}{243} \mathcal{C} \mathcal{H}^2 \right] \Delta^2F^{(3)}(m_K) + \left[ \frac{25}{144} \mathcal{C}^3 - \frac13 \mathcal{C} DF \right] \Delta^2F^{(3)}(m_\eta) \nonumber \\
& & \mbox{} - \mathcal{C} I(m_\pi) - \frac12 \mathcal{C} I(m_K) + \ldots,
\end{eqnarray}

\begin{eqnarray}
\delta g^{\Xi^*\Xi} & = & \left[ \frac{29}{72} \mathcal{C}^3 + \frac{13}{8} \mathcal{C} D^2 + \frac{13}{8} \mathcal{C} F^2 - \frac{13}{4} \mathcal{C} DF + \frac{5}{36} \mathcal{C} D \mathcal{H} - \frac{5}{36} \mathcal{C} F \mathcal{H} + \frac{25}{648} \mathcal{C} \mathcal{H}^2 \right] F^{(1)}(m_\pi) \nonumber \\
& & \mbox{} + \left[ \frac{35}{36} \mathcal{C}^3 - \frac{1}{12} \mathcal{C} D^2 + \frac14 \mathcal{C} F^2 - \frac12 \mathcal{C} DF + \frac{5}{18} \mathcal{C} D \mathcal{H} + \frac{25}{18} \mathcal{C} F \mathcal{H} + \frac{55}{324} \mathcal{C} \mathcal{H}^2 \right] F^{(1)}(m_K) \nonumber \\
& & \mbox{} + \left[ \frac{7}{24} \mathcal{C}^3 - \frac38 \mathcal{C} D^2 + \frac{21}{8} \mathcal{C} F^2 - \frac14 \mathcal{C} DF + \frac{5}{36} \mathcal{C} D \mathcal{H} + \frac{5}{12} \mathcal{C} F \mathcal{H} - \frac{5}{216} \mathcal{C} \mathcal{H}^2 \right] F^{(1)}(m_\eta) \nonumber \\
& & \mbox{} + \left[ \frac18 \mathcal{C}^3 - \frac14 \mathcal{C} D^2 - \frac14 \mathcal{C} F^2 + \frac12 \mathcal{C} DF - \frac{5}{324} \mathcal{C} \mathcal{H}^2 \right] \Delta F^{(2)}(m_\pi) \nonumber \\
& & \mbox{} + \left[ \frac12 \mathcal{C}^3 + \frac23 \mathcal{C} D^2 + \mathcal{C} F^2 + \mathcal{C} DF - \frac{10}{81} \mathcal{C} \mathcal{H}^2 \right] \Delta F^{(2)}(m_K) \nonumber \\
& & \mbox{} + \left[ \frac18 \mathcal{C}^3 + \frac14 \mathcal{C} D^2 - \frac34 \mathcal{C} F^2 + \frac12 \mathcal{C} DF - \frac{5}{108} \mathcal{C} \mathcal{H}^2 \right] \Delta F^{(2)}(m_\eta) \nonumber \\
& & \mbox{} + \left[ \frac{83}{432} \mathcal{C}^3 + \frac{1}{12} \mathcal{C} D^2 + \frac{1}{12} \mathcal{C} F^2 - \frac16 \mathcal{C} DF - \frac{5}{972} \mathcal{C} \mathcal{H}^2 \right] \Delta^2F^{(3)}(m_\pi) \nonumber \\
& & \mbox{} + \left[ \frac{107}{216} \mathcal{C}^3 - \frac29 \mathcal{C} D^2 - \frac13 \mathcal{C} F^2 - \frac13 \mathcal{C} DF - \frac{10}{243} \mathcal{C} \mathcal{H}^2 \right] \Delta^2F^{(3)}(m_K) \nonumber \\
& & \mbox{} + \left[ \frac{25}{144} \mathcal{C}^3 - \frac{1}{12} \mathcal{C} D^2 + \frac14 \mathcal{C} F^2 - \frac16 \mathcal{C} DF - \frac{5}{324} \mathcal{C} \mathcal{H}^2 \right] \Delta^2F^{(3)}(m_\eta) - \mathcal{C} I(m_\pi) - \frac12 \mathcal{C} I(m_K) + \ldots \label{eq:gxsx}
\end{eqnarray}

\begin{eqnarray}
\delta g_A^{\Omega^-\Xi^0} & = & \left[ \frac14 \mathcal{C}^3 + \frac98 \mathcal{C}D^2 - \frac94 \mathcal{C}DF + \frac98 \mathcal{C}F^2 \right] F^{(1)}(m_\pi) \nonumber \\
& & \mbox{} + \left[ \frac76 \mathcal{C}^3 - \frac{1}{12} \mathcal{C} D^2 - \frac52 \mathcal{C}DF + \frac94 \mathcal{C}F^2 + \frac{5}{18} \mathcal{C}D\mathcal{H} + \frac56 \mathcal{C}F\mathcal{H} + \frac{5}{54} \mathcal{C}\mathcal{H}^2 \right] F^{(1)}(m_K) \nonumber \\
& & \mbox{} + \left[ \frac14 \mathcal{C}^3 + \frac18 \mathcal{C}D^2 + \frac34 \mathcal{C}DF + \frac98 \mathcal{C}F^2 + \frac{5}{18} \mathcal{C}D\mathcal{H} + \frac56 \mathcal{C}F\mathcal{H} + \frac{5}{54} \mathcal{C}\mathcal{H}^2 \right] F^{(1)}(m_\eta) \nonumber \\
& & \mbox{} + \frac14 \mathcal{C}^3 \Delta F^{(2)}(m_\pi) + \left[ \frac14 \mathcal{C}^3 + \frac23 \mathcal{C}D^2 + 2 \mathcal{C}DF - \frac{5}{54} \mathcal{C}\mathcal{H}^2 \right] \Delta F^{(2)}(m_K) \nonumber \\
& & \mbox{} + \left[ \frac14 \mathcal{C}^3 - \frac{5}{54} \mathcal{C}\mathcal{H}^2 \right] \Delta F^{(2)}(m_\eta) + \frac18 \mathcal{C}^3 \Delta^2F^{(3)}(m_\pi) \nonumber \\
& & \mbox{} + \left[ \frac{11}{18} \mathcal{C}^3 - \frac29 \mathcal{C}D^2 - \frac23 \mathcal{C}DF - \frac{5}{162} \mathcal{C}\mathcal{H}^2 \right] \Delta^2F^{(3)}(m_K) + \left[ \frac18 \mathcal{C}^3 - \frac{5}{162} \mathcal{C}\mathcal{H}^2 \right] \Delta^2F^{(3)}(m_\eta) \nonumber \\
& & \mbox{} - \frac38 \mathcal{C} I(m_\pi) - \frac34 \mathcal{C} I(m_K) - \frac38 \mathcal{C} I(m_\eta) + \ldots,
\end{eqnarray}

\begin{eqnarray}
\delta g_A^{\Delta^0\Delta^+} & = & \left[ - \frac13 \mathcal{C}^2D - \frac13 \mathcal{C}^2F - \frac{1}{27} \mathcal{C}^2 \mathcal{H} - \frac{52}{243} \mathcal{H}^3\right] F^{(1)}(m_\pi) \nonumber \\
& & \mbox{} + \left[ - \frac23 \mathcal{C}^2F - \frac{5}{27} \mathcal{C}^2 \mathcal{H} - \frac{23}{243} \mathcal{H}^3\right] F^{(1)}(m_K) - \frac{2}{81} \mathcal{H}^3 F^{(1)}(m_\eta) \nonumber \\
& & \mbox{} + \left[ \frac13 \mathcal{C}^2D + \frac13 \mathcal{C}^2F + \frac{5}{27} \mathcal{C}^2 \mathcal{H} \right] \Delta F^{(2)}(m_\pi) + \left[ \frac23 \mathcal{C}^2F + \frac{7}{27} \mathcal{C}^2 \mathcal{H} \right] \Delta F^{(2)}(m_K) \nonumber \\
& & \mbox{} + \left[ - \frac16 \mathcal{C}^2D - \frac16 \mathcal{C}^2F - \frac{19}{162} \mathcal{C}^2 \mathcal{H} \right] \Delta^2F^{(3)}(m_\pi) + \left[ - \frac13 \mathcal{C}^2F - \frac{23}{162} \mathcal{C}^2 \mathcal{H} \right] \Delta^2F^{(3)}(m_K) \nonumber \\
& & \mbox{} + \frac23 \mathcal{H} I(m_\pi) + \frac13 \mathcal{H} I(m_K) + \ldots,
\end{eqnarray}

\begin{eqnarray}
-\sqrt{3} \delta g_A^{\Omega^-{\Xi^*}^0} & = & \left[ \frac18 \mathcal{C}^2 \mathcal{H} + \frac{5}{72} \mathcal{H}^3 \right] F^{(1)}(m_\pi) + \left[ \frac12 \mathcal{C}^2D + \frac32 \mathcal{C}^2F + \frac{5}{12} \mathcal{C}^2 \mathcal{H} + \frac{31}{108} \mathcal{H}^3 \right] F^{(1)}(m_K) \nonumber \\
& & \mbox{} + \left[ - \frac{5}{24} \mathcal{C}^2 \mathcal{H} + \frac{31}{216} \mathcal{H}^3 \right] F^{(1)}(m_\eta) - \frac18 \mathcal{C}^2 \mathcal{H} \Delta F^{(2)}(m_\pi) \nonumber \\
& & \mbox{} + \left[ - \frac12 \mathcal{C}^2D - \frac32 \mathcal{C}^2F - \frac{7}{12} \mathcal{C}^2 \mathcal{H} \right] \Delta F^{(2)}(m_K) + \frac{1}{24} \mathcal{C}^2 \mathcal{H} \Delta F^{(2)}(m_\eta) \nonumber \\
& & \mbox{} + \frac{1}{16} \mathcal{C}^2 \mathcal{H} \Delta^2F^{(3)}(m_\pi) + \left[ \frac14 \mathcal{C}^2D + \frac34 \mathcal{C}^2F + \frac{23}{72} \mathcal{C}^2 \mathcal{H} \right] \Delta^2F^{(3)}(m_K) \nonumber \\
& & \mbox{} + \frac{1}{144} \mathcal{C}^2 \mathcal{H} \Delta^2F^{(3)}(m_\eta) - \frac38 \mathcal{H} I(m_\pi) - \frac34 \mathcal{H} I(m_K) - \frac38 \mathcal{H} I(m_\eta) + \ldots
\end{eqnarray}

\begin{eqnarray}
2\sqrt{3} \delta g_8^{pp} & = & \left[ 3 (-D+3F)(D+F)^2 - 2 \mathcal{C}^2D + 6 \mathcal{C}^2F + \frac{10}{9} \mathcal{C}^2 \mathcal{H} \right] F^{(1)}(m_\pi) \nonumber \\
& & \mbox{} + \left[ - \frac76 D^3 + \frac{13}{2} D^2F - \frac{27}{2} D F^2 + \frac{27}{2} F^3 - \frac52 \mathcal{C}^2D + \frac72 \mathcal{C}^2F \right] F^{(1)}(m_K) \nonumber \\
& & \mbox{} + \frac13 (-D+3F)^3 F^{(1)}(m_\eta) + \left[ - 2 \mathcal{C}^2D + 6 \mathcal{C}^2F + \frac{10}{9} \mathcal{C}^2 \mathcal{H} \right] \Delta F^{(2)}(m_\pi) \nonumber \\
& & \mbox{} + \left[ - \frac32 \mathcal{C}^2D + \frac52 \mathcal{C}^2F \right] \Delta F^{(2)}(m_K) + \left[ - \mathcal{C}^2D + 3
\mathcal{C}^2F + \frac59 \mathcal{C}^2 \mathcal{H} \right] \Delta^2F^{(3)}(m_\pi) \nonumber \\
& & \mbox{} + \left[ - \frac{7}{12} \mathcal{C}^2D + \frac{13}{12} \mathcal{C}^2F \right] \Delta^2F^{(3)}(m_K) - \frac32 (-D+3F) I(m_K) + \ldots. \label{eq:a8}
\end{eqnarray}

\subsection{Perturbative flavor symmetry breaking contributions}

The full expressions containing explicit $SU(3)$ symmetry breaking to the axial couplings are provided here, evaluated at $N_c=3$, for the processes under consideration. They read,
\begin{equation}
\sqrt{3} \delta g_A^{np} = \frac56 z_{1,8} + \frac16 z_{2,8} + \frac{5}{18} z_{4,8} + \frac13 z_{1,10+\overline{10}} + \frac{2}{15} z_{1,27} + \frac{2}{15} z_{2,27} + \frac{1}{30} z_{3,27} + \frac{1}{90} z_{5,27},
\end{equation}

\begin{equation}
\sqrt{2} \delta g_A^{\Sigma^ \pm \Lambda} = \frac13 z_{1,8} + \frac19 z_{4,8} + \frac13 z_{2,10+\overline{10}} - \frac{2}{15} z_{1,27} - \frac{2}{15} z_{2,27} - \frac{1}{30} z_{3,27} - \frac{1}{90} z_{5,27},
\end{equation}

\begin{eqnarray}
\sqrt{2} \delta g_A^{\Lambda p} & = & \frac14 z_{1,8} + \frac{1}{12} z_{2,8} + \frac{1}{12} z_{4,8} - \frac34 z_{1,8_A} - \frac14 z_{2,8_A} - \frac14 z_{4,8_A} - \frac16 z_{1,10+\overline{10}} - \frac16 z_{2,10+\overline{10}} - \frac15 z_{1,27} \nonumber \\
& & \mbox{} - \frac15 z_{2,27} - \frac{1}{20} z_{3,27} - \frac{1}{60} z_{5,27},
\end{eqnarray}

\begin{eqnarray}
\sqrt{3} \delta g_A^{\Sigma^- n} & = & - \frac{1}{12} z_{1,8} + \frac{1}{12} z_{2,8} - \frac{1}{36} z_{4,8} + \frac14 z_{1,8_A} - \frac14 z_{2,8_A} + \frac{1}{12} z_{4,8_A} + \frac16 z_{1,10+\overline{10}} + \frac16 z_{2,10+\overline{10}} \nonumber \\
& & \mbox{} - \frac{1}{15} z_{1,27} - \frac{1}{15} z_{2,27} - \frac{1}{60} z_{3,27} - \frac{1}{180} z_{5,27},
\end{eqnarray}

\begin{eqnarray}
\sqrt{2} \delta g_A^{\Xi^- \Lambda} & = & - \frac{1}{12} z_{1,8} - \frac{1}{12} z_{2,8} - \frac{1}{36} z_{4,8} + \frac14 z_{1,8_A} + \frac14 z_{2,8_A} + \frac{1}{12} z_{4,8_A} + \frac16 z_{1,10+\overline{10}} + \frac16 z_{2,10+\overline{10}} \nonumber \\
& & \mbox{} - \frac15 z_{1,27} - \frac15 z_{2,27} - \frac{1}{20} z_{3,27} - \frac{1}{60} z_{5,27},
\end{eqnarray}

\begin{eqnarray}
\sqrt{6} \delta g_A^{\Xi^- \Sigma^0} & = & - \frac{5}{12} z_{1,8} - \frac{1}{12} z_{2,8} - \frac{5}{36} z_{4,8} + \frac54 z_{1,8_A} + \frac14 z_{2,8_A} + \frac{5}{12} z_{4,8_A} - \frac16 z_{1,10+\overline{10}} - \frac16 z_{2,10+\overline{10}} \nonumber \\
& & \mbox{} - \frac{1}{15} z_{1,27} - \frac{1}{15} z_{2,27} - \frac{1}{60} z_{3,27} - \frac{1}{180} z_{5,27},
\end{eqnarray}

\begin{eqnarray}
\sqrt{3} \delta g_A^{\Xi^0 \Sigma^+} & = & - \frac{5}{12} z_{1,8} - \frac{1}{12} z_{2,8} - \frac{5}{36} z_{4,8} + \frac54 z_{1,8_A} + \frac14 z_{2,8_A} + \frac{5}{12} z_{4,8_A} - \frac16 z_{1,10+\overline{10}} - \frac16 z_{2,10+\overline{10}} \nonumber \\
& & \mbox{} - \frac{1}{15} z_{1,27} - \frac{1}{15} z_{2,27} - \frac{1}{60} z_{3,27} - \frac{1}{180} z_{5,27},
\end{eqnarray}

\begin{equation}
\sqrt{3} \delta g^{\Delta N} = - z_{1,8} + z_{3,8} - \frac12 z_{5,8} - \frac12 z_{1,10+\overline{10}} - \frac12 z_{2,10+\overline{10}} - \frac14 z_{3,10+\overline{10}} - \frac12 z_{4,10+\overline{10}} - \frac{1}{10} z_{1,27} + \frac{1}{10} z_{4,27} - \frac{1}{12} z_{5,27},
\end{equation}

\begin{equation}
\sqrt{3} \delta g^{\Sigma^* \Lambda} = - z_{1,8} + z_{3,8} - \frac12 z_{5,8} + \frac25 z_{1,27} - \frac25 z_{4,27} + \frac13 z_{5,27},
\end{equation}

\begin{equation}
\sqrt{3} \delta g^{\Sigma^* \Sigma} = - z_{1,8} + z_{3,8} - \frac12 z_{5,8} + z_{1,10+\overline{10}} + z_{2,10+\overline{10}} + \frac12 z_{3,10+\overline{10}} + z_{4,10+\overline{10}} - \frac35 z_{1,27} + \frac35 z_{4,27} - \frac12 z_{5,27},
\end{equation}

\begin{equation}
\sqrt{3} \delta g^{\Xi^* \Xi} = - z_{1,8} + z_{3,8} - \frac12 z_{5,8} + z_{1,10+\overline{10}} + z_{2,10+\overline{10}} + \frac12 z_{3,10+\overline{10}} + z_{4,10+\overline{10}} + \frac25 z_{1,27} - \frac25 z_{4,27} + \frac13 z_{5,27},
\end{equation}

\begin{equation}
\sqrt{3} \delta g_A^{\Omega^- \Xi^0} = \frac12 z_{1,8} - \frac12 z_{3,8} - \frac32 z_{1,8_A}+ \frac12 z_{1,10+\overline{10}} + \frac12 z_{2,10+\overline{10}} + \frac14 z_{3,10+\overline{10}} + \frac{3}{10} z_{1,271} - \frac{3}{10} z_{4,27} + \frac14 z_{5,27},
\end{equation}

\begin{equation}
\sqrt{3} \delta g_A^{\Delta^0 \Delta^+} = z_{1,8} + z_{2,8} + \frac53 z_{4,8} + \frac25 z_{1,27} + \frac45 z_{2,27} + \frac15 z_{3,27} + \frac13 z_{5,27},
\end{equation}

\begin{equation}
 \delta g_A^{\Omega^- {\Xi^*}^0} = - \frac14 z_{1,8} - \frac14 z_{2,8} - \frac{5}{12} z_{4,8} + \frac34 z_{1,8_A} + \frac34 z_{2,8_A} + \frac54 z_{4,8_A} - \frac35 z_{1,27} - \frac65 z_{2,27} - \frac{3}{10} z_{3,27} - \frac12 z_{5,27},
\end{equation}

\begin{equation}
(\delta g_A^{pp})_8 = \frac12 z_{11} + \frac{1}{12} z_{2,1} - \frac{1}{12} z_{1,8} - \frac{1}{12} z_{2,8} - \frac{1}{36} z_{4,8} + \frac{1}{20} z_{1,27} + \frac{1}{20} z_{2,27} + \frac{1}{80} z_{3,27} + \frac{1}{240} z_{5,27}.
\end{equation}

\section{\label{app:lecs}Further remarks about low-energy constants}

The calculational scheme presented here faces some challenging aspects. The main idea is to include the dependence on $m/\Delta$ of the loop integrals, evaluated at its physical value. An alternative approach found in the literature is not to include the intermediate spin-3/2 baryons explicitly in the loops; instead, their effects are accounted for in the low-energy constants of the effective Lagrangian.

Analyses of low-energy constants have been extensively presented in the literature with the introduction of Lagrangians $\mathcal{L}_{\pi N}^{(2,3)}$ and even $\mathcal{L}_{\pi N}^{(4)}$. Important results can be traced back to Refs.~\cite{krause,b3}, and more recently,  \cite{heo}, to name but a few.

In the present analysis, the effects of the low-energy constants were accounted for with the introduction of explicit symmetry breaking terms in the axial vector current operator, parameterized in terms of the factors $z_{j,\mathrm{rep}}$ introduced in Eqs.~(\ref{eq:sb1})-(\ref{eq:sb10}) and fully listed in Appendix \ref{app:sb}. Some tree-diagram contributions are needed as counterterms for the divergent parts of the loop integrals.

\begin{figure}[ht]
\scalebox{0.3}{\includegraphics{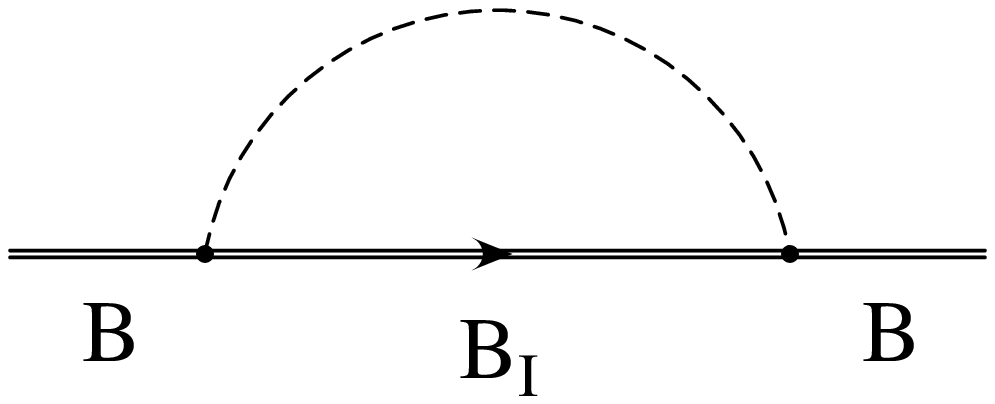}}
\caption{\label{fig:wave}One-loop wave function renormalization graph.}
\end{figure}

It would be interesting to outline how the operator parameters $z_{j,\mathrm{rep}}$ can be related with those low-energy constants. To start with, it is illustrative to analyze the wave function renormalization diagram depicted in Fig.~\ref{fig:wave}.

The loop graph can be written as \cite{rfm00}
\begin{equation}
iG_{BB^\prime} = \sum_{j,k,b,B_I} \left[A^{kb}\right]_{B^\prime B_I} \left[A^{jb}\right]_{B_IB} \int \frac{d^4k}{(2\pi)^4} \frac{(\mathbf{k}^k)(-\mathbf{k}^j)}{(k^2-m_b^2)[(k+p)\cdot v-(M_I-M)+i \epsilon]},
\end{equation}
where $b$ labels mesons with momentum $\mathbf{k}$ and $B_I$ denotes the intermediate baryon in the loop. The wave function renormalization correction is
\begin{equation}
Z_{BB^\prime} = \delta_{BB^\prime} + z_{BB^\prime},
\end{equation}
where
\begin{equation}
z_{BB^\prime} = \frac{\partial G_{BB^\prime}}{\partial (v\cdot p)}.
\end{equation}

Now, the replacement of summation over intermediate baryon states with matrix multiplication yields
\begin{equation}
z_{BB^\prime} = - A^{ja}A^{jb} \Pi_{(1)}^{ab},
\end{equation}
where the symmetric tensor $\Pi_{(1)}^{ab}$ in given in Eq.~(\ref{eq:ibp}). Thus, $z_{BB^\prime}$ can be decomposed into $1$, $8$ and $27$ $SU(3)$ flavor components. The operator reductions required are listed below. \\

1 representation

\begin{equation}
G^{ia} G^{ia} = \frac{3}{16} N_c(N_c+2N_f) - \frac{N_f+2}{4N_f} J^2,
\end{equation}

\begin{equation}
G^{ia} \mathcal{D}_2^{ia} + \mathcal{D}_2^{ia}G^{ia} = \frac{(N_c+N_f)(N_f-1)}{N_f} J^2,
\end{equation}

\begin{equation}
G^{ia} \mathcal{D}_3^{ia} + \mathcal{D}_3^{ia}G^{ia} = \frac12 (N_c+2N_f-2)(N_c+2) J^2 + \frac{N_f-2}{N_f} \{J^2,J^2\},
\end{equation}

\begin{equation}
G^{ia} \mathcal{O}_3^{ia} + \mathcal{O}_3^{ia}G^{ia} = \frac34 N_c(N_c+2N_f) + \frac12 [N_c(N_c+2N_f)-6N_f] J^2 - \{J^2,J^2\},
\end{equation}

\begin{equation}
\mathcal{D}_2^{ia} \mathcal{D}_2^{ia} = \frac{N_c(N_c+2N_f)(N_f-2)}{4N_f} J^2 + \frac12 \{J^2,J^2\},
\end{equation}

\begin{equation}
\mathcal{D}_2^{ia} \mathcal{D}_3^{ia} + \mathcal{D}_3^{ia}\mathcal{D}_2^{ia} = \frac{2(N_c+N_f)(N_f-1)}{N_f} \{J^2,J^2\},
\end{equation}

\begin{equation}
\mathcal{D}_2^{ia} \mathcal{O}_3^{ia} + \mathcal{O}_3^{ia}\mathcal{D}_2^{ia} = 0,
\end{equation}

\begin{equation}
\mathcal{D}_3^{ia} \mathcal{D}_3^{ia} = \frac12 (N_c+2N_f-2)(N_c+2) \{J^2,J^2\} + \frac{N_f-2}{N_f} \{J^2,\{J^2,J^2\}\},
\end{equation}

\begin{equation}
\mathcal{D}_3^{ia} \mathcal{O}_3^{ia} + \mathcal{O}_3^{ia}\mathcal{D}_3^{ia} = 0,
\end{equation}

\begin{equation}
\mathcal{O}_3^{ia} \mathcal{O}_3^{ia} = \frac34 N_c(N_c+2N_f) + [2N_c(N_c+2N_f)-3N_f] J^2 + \frac14 [N_c(N_c+2N_f)-10N_f-6] \{J^2,J^2\} - \frac12 \{J^2,\{J^2,J^2\}\}.
\end{equation}

8 representation

\begin{equation}
d^{ab8} G^{ia} G^{ib} = \frac38 (N_c+N_f) T^8 - \frac{N_f+4}{4N_f} \{J^i,G^{i8}\},
\end{equation}

\begin{equation}
d^{ab8} (G^{ia} \mathcal{D}_2^{ib} + \mathcal{D}_2^{ia} G^{ib}) = \frac{(N_c+N_f)(N_f-2)}{2N_f} \{J^i,G^{i8}\} + \frac{N_f-2}{2N_f} \{J^2,T^8\},
\end{equation}

\begin{equation}
d^{ab8} (G^{ia} \mathcal{D}_3^{ib} + \mathcal{D}_3^{ia} G^{ib}) = (N_f-2) \{J^i,G^{i8}\} + \frac12 (N_c+N_f) \{J^2,T^8\} + \frac{N_f-4}{N_f} \{J^2,\{J^i,G^{i8}\}\},
\end{equation}

\begin{equation}
d^{ab8} (G^{ia} \mathcal{O}_3^{ib} + \mathcal{O}_3^{ia} G^{ib}) = \frac32 (N_c+N_f) T^8 - \frac32 N_f \{J^i,G^{i8}\} + \frac12 (N_c+N_f) \{J^2,T^8\} - \{J^2,\{J^i,G^{i8}\}\},
\end{equation}

\begin{equation}
d^{ab8} \mathcal{D}_2^{ia} \mathcal{D}_2^{ib} = \frac{(N_c+N_f)(N_f-4)}{4N_f} \{J^2,T^8\} + \frac12 \{J^2,\{J^i,G^{i8}\}\},
\end{equation}

\begin{equation}
d^{ab8}(\mathcal{D}_2^{ia} \mathcal{D}_3^{ib} + \mathcal{D}_3^{ia} \mathcal{D}_2^{ib}) = \frac{(N_c+N_f)(N_f-2)}{N_f} \{J^2,\{J^i,G^{i8}\}\} + \frac{N_f-2}{N_f} \{J^2,\{J^2,T^8\}\},
\end{equation}

\begin{equation}
d^{ab8} (\mathcal{D}_2^{ia} \mathcal{O}_3^{ib} + \mathcal{O}_3^{ia} \mathcal{D}_2^{ib}) = 0,
\end{equation}

\begin{equation}
d^{ab8} \mathcal{D}_3^{ia} \mathcal{D}_3^{ib} = (N_f-2) \{J^2,\{J^i,G^{i8}\}\} + \frac12 (N_c+N_f) \{J^2,\{J^2,T^8\}\} + \frac{N_f-4}{N_f} \{J^2,\{J^2,\{J^i,G^{i8}\}\}\},
\end{equation}

\begin{equation}
d^{ab8} (\mathcal{D}_3^{ia} \mathcal{O}_3^{ib} + \mathcal{O}_3^{ia} \mathcal{D}_3^{ib}) = 0,
\end{equation}

\begin{eqnarray}
d^{ab8} \mathcal{O}_3^{ia} \mathcal{O}_3^{ib} & = & \frac32 (N_c+N_f) T^8 - \frac32 N_f \{J^i,G^{i8}\} + 2 (N_c+N_f) \{J^2,T^8\} - \frac14 (5N_f+6) \{J^2,\{J^i,G^{i8}\}\} \nonumber \\
& & \mbox{} + \frac14 (N_c+N_f) \{J^2,\{J^2,T^8\}\} - \frac12 \{J^2,\{J^2,\{J^i,G^{i8}\}\}\}.
\end{eqnarray}

27 representation

\begin{equation}
G^{i8} G^{i8} = \frac12 \{G^{r8},G^{r8}\},
\end{equation}

\begin{equation}
G^{i8} \mathcal{D}_2^{i8} + \mathcal{D}_2^{i8} G^{i8} = \frac12 \{T^8,\{J^r,G^{r8}\}\},
\end{equation}

\begin{equation}
G^{i8} \mathcal{D}_3^{i8} + \mathcal{D}_3^{i8} G^{i8} = \{\{J^i,G^{i8}\},\{J^j,G^{j8}\}\},
\end{equation}

\begin{equation}
G^{i8} \mathcal{O}_3^{i8} + \mathcal{O}_3^{i8} G^{i8} = - \frac{2}{N_f} \delta^{88} J^2 + 2 \{G^{r8},G^{r8}\} - d^{88e} \{J^r,G^{re}\} + \{J^2,\{G^{r8},G^{r8}\}\} - \frac12 \{\{J^i,G^{i8}\},\{J^j,G^{j8}\}\},
\end{equation}

\begin{equation}
\mathcal{D}_2^{i8} \mathcal{D}_2^{i8} = \frac14 \{J^2,\{T^8,T^8\}\},
\end{equation}

\begin{equation}
\mathcal{D}_2^{i8} \mathcal{D}_3^{i8} + \mathcal{D}_3^{i8} \mathcal{D}_2^{i8} = \{J^2,\{T^8,\{J^r,G^{r8}\}\}\},
\end{equation}

\begin{equation}
\mathcal{D}_2^{i8} \mathcal{O}_3^{i8} + \mathcal{O}_3^{i8} \mathcal{D}_2^{i8} = 0,
\end{equation}

\begin{equation}
\mathcal{D}_3^{i8} \mathcal{D}_3^{i8} = \{J^2,\{\{J^i,G^{i8}\},\{J^j,G^{j8}\}\}\},
\end{equation}

\begin{equation}
\mathcal{D}_3^{i8} \mathcal{O}_3^{i8} + \mathcal{O}_3^{i8} \mathcal{D}_3^{i8} = 0,
\end{equation}

\begin{eqnarray}
\mathcal{O}_3^{i8} \mathcal{O}_3^{i8} & = & - \frac{2}{N_f} \delta^{88}J^2 + 2 \{G^{r8},G^{r8}\} - d^{88e} \{J^r,G^{re}\} + 3
 \{J^2,\{G^{r8},G^{r8}\}\} - \frac12 \{\{J^i,G^{i8}\},\{J^j,G^{j8}\}\} \nonumber \\
& & \mbox{} - d^{88e} \{J^2,\{J^r,G^{re}\}\} - \frac{2}{N_f} \delta^{88} \{J^2,J^2\} - \frac14 \{J^2,\{\{J^i,G^{i8}\},\{J^j,G^{j8}\}\}\} \nonumber \\
& & \mbox{} + \frac12 \{J^2,\{J^2,\{G^{r8},G^{r8}\}\}\}.
\end{eqnarray}

Notice that the operator structures $A^{ia}A^{ia}$ are precisely those used in the construction of the $1/N_c$ expansion of the baryon mass operator, whereas $d^{ab8} A^{ia}A^{ib}$ and $A^{i8}A^{i8}$ are closely related to first- and second-order flavor symmetry breaking in that baryon mass operator \cite{rfm24}.

The matrix elements of the operator $z_{BB^\prime}$ can thus be organized as $[z]_{BB^\prime}=[z_1]_{BB^\prime}+[z_8]_{BB^\prime}+[z_{27}]_{BB^\prime}$. For the nucleon $N$, for instance, it is found that
\begin{eqnarray}
-[z_1]_N & = & \left[ \frac{57}{32} a_1^2 + \frac38 a_1b_2 + \frac{9}{16} a_1b_3 + \frac{15}{16} a_1c_3 + \frac{3}{32} b_2^2 + \frac18b_2b_3 + \frac{3}{32} b_3^2 + \frac{15}{64} c_3^2 \right] F^{(1)}(m_\pi) \nonumber \\
& & \mbox{} + \left[ \frac{19}{8} a_1^2 + \frac12 a_1b_2 + \frac34 a_1b_3 + \frac54 a_1c_3 + \frac18 b_2^2 + \frac16b_2b_3 + \frac18b_3^2 + \frac{5}{16} c_3^2 \right] F^{(1)}(m_K) \nonumber \\
& & \mbox{} + \left[ \frac{19}{32} a_1^2 + \frac18 a_1b_2 + \frac{3}{16} a_1b_3 + \frac{5}{16} a_1c_3 + \frac{1}{32} b_2^2 + \frac{1}{24} b_2b_3 + \frac{1}{32} b_3^2 + \frac{5}{64} c_3^2 \right] F^{(1)}(m_\eta),
\end{eqnarray}

\begin{eqnarray}
- [z_8]_N & = & \left[ \frac{141}{80} a_1^2 + \frac{9}{40} a_1b_2 + \frac{19}{40} a_1b_3 + \frac{21}{20} a_1c_3 - \frac{3}{80} b_2^2 + \frac{3}{40} b_2b_3 + \frac{19}{240} b_3^2 + \frac{21}{80} c_3^2 \right] F^{(1)}(m_\pi) \nonumber \\
& & \mbox{} + \left[ - \frac{47}{40} a_1^2 - \frac{3}{20} a_1b_2 - \frac{19}{60} a_1b_3 - \frac{7}{10} a_1c_3 + \frac{1}{40} b_2^2 - \frac{1}{20} b_2b_3 - \frac{19}{360} b_3^2 - \frac{7}{40} c_3^2 \right] F^{(1)}(m_K) \nonumber \\
& & \mbox{} + \left[ - \frac{47}{80} a_1^2 - \frac{3}{40} a_1b_2 - \frac{19}{120} a_1b_3 - \frac{7}{20} a_1c_3 + \frac{1}{80} b_2^2 - \frac{1}{40} b_2b_3 - \frac{19}{720} b_3^2 - \frac{7}{80} c_3^2 \right] F^{(1)}(m_\eta), \nonumber \\
\end{eqnarray}

\begin{eqnarray}
-[z_{27}]_N & = & \left[ \frac{3}{160} a_1^2 + \frac{1}{40} a_1b_2 + \frac{1}{240} a_1b_3 + \frac{1}{80} a_1c_3 + \frac{1}{160} b_2^2 + \frac{1}{120} b_2b_3 + \frac{1}{1440} b_3^2 + \frac{1}{320} c_3^2 \right] F^{(1)}(m_\pi) \nonumber \\
& & \mbox{} + \left[ - \frac{3}{40} a_1^2 - \frac{1}{10} a_1b_2 - \frac{1}{60} a_1b_3 - \frac{1}{20} a_1c_3 - \frac{1}{40} b_2^2 - \frac{1}{30} b_2b_3 - \frac{1}{360} b_3^2 - \frac{1}{80} c_3^2 \right] F^{(1)}(m_K) \nonumber \\
& & \mbox{} + \left[ \frac{9}{160} a_1^2 + \frac{3}{40} a_1b_2 + \frac{1}{80} a_1b_3 + \frac{3}{80} a_1c_3+ \frac{3}{160} b_2^2 + \frac{1}{40} b_2b_3 + \frac{1}{480} b_3^2 + \frac{3}{320} c_3^2 \right] F^{(1)}(m_\eta), \nonumber \\
\end{eqnarray}
or in terms of the $SU(3)$ coupling constants they can be written as
\begin{eqnarray}
-[z_1]_N & = & \left[ \frac{15}{8} D^2 + \frac{27}{8} F^2 + \frac{15}{16} \mathcal{C}^2 \right] F^{(1)}(m_\pi) + \left[ \frac52 D^2 + \frac92 F^2 + \frac54 \mathcal{C}^2 \right] F^{(1)}(m_K) \nonumber \\
& & \mbox{} + \left[ \frac58 D^2 + \frac98 F^2 + \frac{5}{16} \mathcal{C}^2 \right] F^{(1)}(m_\eta),
\end{eqnarray}

\begin{eqnarray}
- [z_8]_N & = & \left[\frac{9}{20} D^2 + \frac92 DF - \frac{27}{20} F^2 + \frac{21}{20} \mathcal{C}^2 \right] F^{(1)}(m_\pi) + \left[ - \frac{3}{10} D^2 - 3 DF + \frac{9}{10} F^2 - \frac{7}{10} \mathcal{C}^2 \right] F^{(1)}(m_K) \nonumber \\
& & \mbox{} + \left[ - \frac{3}{20} D^2 - \frac32 DF + \frac{9}{20} F^2 - \frac{7}{20} \mathcal{C}^2 \right] F^{(1)}(m_\eta) 
\end{eqnarray}

\begin{eqnarray}
- [z_{27}]_N & = & \left[ - \frac{3}{40} D^2 + \frac{9}{40} F^2 + \frac{1}{80} \mathcal{C}^2\right] F^{(1)}(m_\pi) + \left[ \frac{3}{10} D^2 - \frac{9}{10} F^2 - \frac{1}{20} \mathcal{C}^2 \right] F^{(1)}(m_K) \nonumber \\
& & \mbox{} + \left[ - \frac{9}{40} D^2 + \frac{27}{40} F^2 + \frac{3}{80} \mathcal{C}^2 \right] F^{(1)}(m_\eta).
\end{eqnarray}

In the limits $m_u=m_d=0$, $m_\eta^2 \to \frac43 m_K^2$ and $\Delta \to 0$ and omitting the divergent terms, $Z_N=1+z_N$ is in full agreement with the corresponding $Z_N=1+\lambda_N (m_K^2/(16\pi f^2) \log(m_K^2/\mu^2)$, given in Eq.~(52) of Ref.~\cite{jen91}. The agreement is also observed for the other baryons.

The appropriate handling of the divergent terms in $iG_{BB^\prime}$ for the nucleon (without considering explicit decuplet degrees of freedom)  requires the introduction of three contact terms $c_1$ and $B_{20}$, and $B_{15}$  of order $p^2$ and $p^3$, respectively \cite{b3}.

The chiral Lagrangian extended to three flavors has been recently revisited in Ref.~\cite{heo}, where a systematic analysis of all low-energy constants (23 parameters in total) that contribute to this chiral order is given. Applications to the axial-vector and pseudoscalar currents in the baryon octet and decuplet fields at next-to-leading order is also presented in detail in Ref.~\cite{heo}.
Results specialized to the axial-vector coupling for the beta decay of octet baryons and the hadronic coupling constants for decuplet baryons, decomposed into low-energy parameters, are listed in Tables 1 and 2 of this reference. These parameters can be related to the $z_{j,\mathrm{dim}}$ introduced here in the following manner,
\begin{equation}
F_{1} = \frac{\varepsilon_0}{\varepsilon_8} F_{0} - \frac{1}{12 \sqrt{3} \varepsilon_8} (6 z_{1,8} + 3 z_{2,8} + 2 z_{4,8} + 3 z_{1,8_A} - 3 z_{2,8_A} + z_{4,8_A} + 3 z_{1,10+\overline{10}} - 4 z_{2,10+\overline{10}}),
\end{equation}
\begin{equation}
F_{2} = - \frac{\varepsilon_0}{\varepsilon_8} F_{0} - \frac{1}{12 \sqrt{3} \varepsilon_8} (6 z_{1,8} + 3 z_{2,8} + 2 z_{4,8} - 15 z_{1,8_A} - 3 z_{2,8_A} - 5 z_{4,8_A} - 3 z_{1,10+\overline{10}} - 16 z_{2,10+\overline{10}}),
\end{equation}
\begin{equation}
F_{3} = - \frac{4 \varepsilon_0 - \varepsilon_8}{3 \varepsilon_8}F_{0} - \frac{1}{90 \sqrt{3}\varepsilon_8} (45 z_{1,8} + 15 z_{4,8} - 135 z_{1,8_A} - 45 z_{4,8_A} + 12 z_{1,27} + 12 z_{2,27} + 3 z_{3,27} + z_{5,27}),
\end{equation}
\begin{equation}
F_{5} = \frac43 \frac{\varepsilon_0 - \varepsilon_8}{\varepsilon_8} F_{0} - \frac{1}{30 \sqrt{3}\varepsilon_8} (30 z_{1,8} + 10 z_{4,8} + 30 z_{2,10+\overline{10}} - 12 z_{1,27} - 12 z_{2,27} - 3 z_{3,27} - z_{5,27}),
\end{equation}
\begin{eqnarray}
F_{6} & = & \frac23 \frac{\varepsilon_0 - \varepsilon_8}{\varepsilon_8} F_{0} + \frac{1}{18\sqrt{3} \varepsilon_0} (15 z_{1,8_A} + 3 z_{2,8_A} + 5 z_{4,8_A} - 2 z_{2,10+\overline{10}}) \nonumber \\
& & \mbox{} + \frac{1}{18 \sqrt{3} \varepsilon_8} (6 z_{1,8} + 3 z_{2,8} + 2 z_{4,8} - 15 z_{1,8_A} - 3 z_{2,8_A} - 5 z_{4,8_A} - 3 z_{1,10+\overline{10}} - 16 z_{2,10+\overline{10}}),
\end{eqnarray}
\begin{equation}
C_{3} = - \frac13 C_{2} + \frac{1}{12 \sqrt{3} \varepsilon_8} (6 z_{1,10+\overline{10}} + 6 z_{2,10+\overline{10}} + 3 z_{3,10+\overline{10}} + 6 z_{4,10+\overline{10}} + 6 z_{1,27} - 6 z_{4,27} + 5 z_{5,27}),
\end{equation}
\begin{equation}
C_{4} = - \frac23 C_{2} - \frac{1}{12 \sqrt{3} \varepsilon_8} (6 z_{1,10+\overline{10}} + 6 z_{2,10+\overline{10}} + 3 z_{3,10+\overline{10}} + 6 z_{4,10+\overline{10}} - 6 z_{1,27} + 6 z_{4,27} - 5 z_{5,27}),
\end{equation}
and
\begin{equation}
C_{5} = - C_{0} - \frac{\varepsilon_0 + \varepsilon_8}{\varepsilon_0 - \varepsilon_8} C_{2} - \frac{1}{20 \sqrt{3} (\varepsilon_0 - \varepsilon_8)} (30 z_{1,8} - 30 z_{3,8} + 15 z_{5,8} - 12 z_{1,27} + 12 z_{4,27} - 10 z_{5,27}).
\end{equation}
where $F_m$ and $C_n$ are the low-energy constants to chiral order $p^3$ introduced in Ref.~\cite{heo} and $\varepsilon_0 \simeq 2m_K^2+m_\pi^2$ and $\varepsilon_8 \simeq 2(m_K^2-m_\pi^2)$.


\begin{thebibliography}{99}

\bibitem{jm255}
E.~Jenkins and A.~V.~Manohar,
``Baryon chiral perturbation theory using a heavy fermion Lagrangian,''
Phys.\ Lett.\ B {\bf 255}, 558 (1991).

\bibitem{jm259}
E.~Jenkins and A.~V.~Manohar,
``Chiral corrections to the baryon axial currents,''
Phys.\ Lett.\ B {\bf 259}, 353 (1991).

\bibitem{jen91}
E.~E.~Jenkins and A.~V.~Manohar,
``Baryon chiral perturbation theory,'' UCSD-PTH-91-30.

\bibitem{tHooft}
G.~'t Hooft,
``A two-dimensional model for mesons,''
Nucl.\ Phys.\ B \textbf{75}, 461 (1974).

\bibitem{ven}
G.~Veneziano,
``Some aspects of a unified approach to gauge, dual and Gribov theories,''
Nucl.\ Phys.\ B \textbf{117}, 519 (1976).

\bibitem{witten}
E.~Witten,
``Baryons in the $1/N$ expansion,''
Nucl.\ Phys.\ B {\bf 160}, 57 (1979).

\bibitem{dm1}
R.~F.~Dashen and A.~V.~Manohar,
\lq\lq Baryon-pion couplings from large-$N_c$ QCD,"
Phys.\ Lett.\ B \textbf{315}, 425 (1993).

\bibitem{djm94}
R.~F.~Dashen, E.~E.~Jenkins and A.~V.~Manohar,
``The $1/N_c$ expansion for baryons,''
Phys.\ Rev.\ D \textbf{49}, 4713 (1994);
[erratum: Phys.\ Rev.\ D \textbf{51}, 2489 (1995).]

\bibitem{djm95}
R.~F.~Dashen, E.~Jenkins and A.~V.~Manohar,
``Spin-flavor structure of large $N_c$ baryons,''
Phys.\ Rev.\ D {\bf 51}, 3697 (1995).

\bibitem{jen96}
E.~Jenkins,
``Chiral Lagrangian for baryons in the $1/N_c$ expansion,''
Phys.\ Rev.\ D {\bf 53}, 2625 (1996).

\bibitem{weise}
Y.~s.~Oh and W.~Weise,
``Baryon masses in large $N_c$ chiral perturbation theory,''
Eur.\ Phys.\ J.\ A \textbf{4}, 363 (1999).

\bibitem{rfm00}
R.~Flores-Mendieta, C.~P.~Hofmann, E.~Jenkins and A.\ V.\ Manohar,
``Structure of large-$N_c$ cancellations in baryon chiral perturbation theory,''
Phys.\ Rev.\ D {\bf 62}, 034001 (2000).

\bibitem{rfm06}
R.~Flores-Mendieta and C.~P.~Hofmann,
``Renormalization of the baryon axial vector current in large-$N_c$ chiral perturbation theory,''
Phys.\ Rev.\ D \textbf{74}, 094001 (2006).

\bibitem{rfm12}
R.~Flores-Mendieta, M.~A.~Hernandez-Ruiz and C.~P.~Hofmann,
``Renormalization of the baryon axial vector current in large-$N_c$ chiral perturbation theory: Effects of the decuplet-octet mass difference and flavor symmetry breaking,''
Phys.\ Rev.\ D \textbf{86}, 094041 (2012).

\bibitem{fg0}
I.~P.~Fernando and J.~L.~Goity,
``Baryon chiral perturbation theory combined with the $1/N_c$ expansion in SU(3): Framework,''
Phys.\ Rev.\ D \textbf{97}, 054010  (2018).

\bibitem{rfm21}
R.~Flores-Mendieta, C.~I.~Garcia and J.~Hernandez,
``Baryon axial vector current in large-$N_c$ chiral perturbation theory: Complete analysis for $N_c=3$,''
Phys.\ Rev.\ D \textbf{103}, 094032 (2021).

\bibitem{fmg}
R.~Flores-Mendieta and J.~L.~Goity,
``Baryon vector current in the chiral and $1/N_c$ expansions,''
Phys. Rev. D \textbf{90}, 114008 (2014).

\bibitem{fg}
I.~P.~Fernando and J.~L.~Goity,
``$SU(3)$ vector currents in baryon chiral perturbation theory combined with the $1/N_c$ expansion,''
Phys.\ Rev.\ D \textbf{101}, 054026 (2020).

\bibitem{luty}
M.~A.~Luty, J.~March-Russell and M.~J.~White,
``Baryon magnetic moments in a simultaneous expansion in $1/N$ and $m_s$,''
Phys.\ Rev.\ D \textbf{51},2332 (1995).

\bibitem{rfm09}
R.~Flores-Mendieta,
``Baryon magnetic moments in large-$N_c$ chiral perturbation theory,''
Phys.\ Rev.\ D \textbf{80}, 094014 (2009).

\bibitem{rfm14b}
G.~Ahuatzin, R.~Flores-Mendieta, M.~A.~Hernandez-Ruiz, and C.~P.~Hofmann,
``Baryon magnetic moments in large-$N_c$ chiral perturbation theory: Effects of the decuplet-octet mass difference and flavor symmetry breaking,''
Phys.\ Rev.\ D \textbf{89}, 034012 (2014).

\bibitem{rfm21b}
R.~Flores-Mendieta, C.~I.~Garcia, J.~Hernandez and M.~A.~Trejo,
``Baryon magnetic moment in large-$N_c$ chiral perturbation theory: Complete analysis for $N_c=3$,''
Phys.\ Rev.\ D \textbf{104}, 114024 (2021).

\bibitem{dai}
J.~Dai, R.~F.~Dashen, E.~Jenkins and A.~V.~Manohar,
``Flavor symmetry breaking in the $1/N_c$ expansion,''
Phys.\ Rev.\ D {\bf 53}, 273 (1996).

\bibitem{lacour}
A.~Lacour, B.~Kubis and U.~G.~Meissner,
``Hyperon decay form-factors in chiral perturbation theory,''
JHEP \textbf{10}, 083 (2007).

\bibitem{sau1}
U.~Sauerwein, M.~F.~M.~Lutz and R.~G.~E.~Timmermans,
``Axial-vector form factors of the baryon octet and chiral symmetry,''
Phys.\ Rev.\ D \textbf{105}, 054005 (2022).

\bibitem{sau2}
M.~F.~M.~Lutz, U.~Sauerwein and R.~G.~E.~Timmermans,
``On the axial-vector form factor of the nucleon and chiral symmetry,''
Eur.\ Phys.\ J.\ C \textbf{80}, 844 (2020).

\bibitem{b1}
V.~Bernard, N.~Kaiser, T.~S.~H.~Lee and U.~G.~Meissner,
``Threshold pion electroproduction in chiral perturbation theory,''
Phys.\ Rept.\ \textbf{246}, 315 (1994).

\bibitem{b2}
V.~Bernard, H.~W.~Fearing, T.~R.~Hemmert and U.~G.~Meissner,
``The form-factors of the nucleon at small momentum transfer,''
Nucl.\ Phys.\ A \textbf{635}, 121 (1998).
[erratum: Nucl.\ Phys.\ A \textbf{642}, 563 (1998)].

\bibitem{sch}
M.~R.~Schindler, T.~Fuchs, J.~Gegelia and S.~Scherer,
``Axial, induced pseudoscalar, and pion-nucleon form-factors in manifestly Lorentz-invariant chiral perturbation theory,''
Phys.\ Rev.\ C \textbf{75}, 025202 (2007).

\bibitem{fuchs}
T.~Fuchs, J.~Gegelia, G.~Japaridze and S.~Scherer,
``Renormalization of relativistic baryon chiral perturbation theory and power counting,''
Phys.\ Rev.\ D \textbf{68}, 056005 (2003).

\bibitem{geng1}
L.~S.~Geng, J.~Martin Camalich, L.~Alvarez-Ruso and M.~J.~Vicente Vacas,
``Nucleon-to-Delta axial transition form factors in relativistic baryon chiral perturbation theory,''
Phys.\ Rev.\ D \textbf{78}, 014011 (2008).

\bibitem{geng2}
L.~S.~Geng, J.~Martin Camalich and M.~J.~Vicente Vacas,
``SU(3)-breaking corrections to the hyperon vector coupling $f_1(0)$ in covariant baryon chiral perturbation theory,''
Phys.\ Rev.\ D \textbf{79}, 094022 (2009).

\bibitem{suh}
J.~M.~Suh, Y.~S.~Jun and H.~C.~Kim,
``Axial-vector transition form factors of the baryon octet to the baryon decuplet with flavor SU(3) symmetry breaking,''
Phys.\ Rev.\ D \textbf{105}, 114040 (2022).

\bibitem{dahiya}
H.~Dahiya, S.~Dutt, A.~Kumar and M.~Randhawa,
``Axial-vector charges of the spin $\frac{1}{2}^+$ and spin $\frac{3}{2}^+$ light and charmed baryons in the SU(4) chiral quark constituent model,''
Eur.\ Phys.\ J.\ Plus \textbf{138}, 441 (2023).

\bibitem{gu}
D.~Guadagnoli, V.~Lubicz, M.~Papinutto and S.~Simula,
``First lattice QCD study of the $\Sigma^- \to n$ axial and vector form factors with $SU(3)$ breaking corrections,''
Nucl.\ Phys.\ B \textbf{761}, 63 (2007).

\bibitem{pndme}
Y.~C.~Jang \textit{et al.} [Precision Neutron Decay Matrix Elements (PNDME) Collaboration],
``Nucleon isovector axial form factors,''
Phys.\ Rev.\ D \textbf{109}, 014503 (2024).

\bibitem{rqcd}
G.~S.~Bali \textit{et al.} [RQCD Collaboration],
``Nucleon axial structure from lattice QCD,''
JHEP \textbf{05}, 126 (2020).

\bibitem{ale}
C.~Alexandrou \textit{et al.} [Extended Twisted Mass Collaboration]
``Nucleon axial and pseudoscalar form factors from lattice QCD at the physical point,''
Phys.\ Rev.\ D \textbf{103}, 034509 (2021).

\bibitem{pacs}
R.~Tsuji \textit{et al.} [PACS Collaboration],
``Nucleon isovector couplings in $N_f=2+1$ lattice QCD at the physical point,''
Phys.\ Rev.\ D \textbf{106}, 094505 (2022).

\bibitem{dju}
D.~Djukanovic, G.~von Hippel, J.~Koponen, H.~B.~Meyer, K.~Ottnad, T.~Schulz and H.~Wittig,
``Isovector axial form factor of the nucleon from lattice QCD,''
Phys.\ Rev.\ D \textbf{106}, 074503 (2022).

\bibitem{nme}
S.~Park \textit{et al.} [Nucleon Matrix Elements (NME) Collaboration],
``Precision nucleon charges and form factors using (2+1)-flavor lattice QCD,''
Phys.\ Rev.\ D \textbf{105}, 054505 (2022).

\bibitem{flag}
Y.~Aoki \textit{et al.} [Flavour Lattice Averaging Group (FLAG)],
``FLAG Review 2024,''
[arXiv:2411.04268 [hep-lat]].

\bibitem{banda2}
V.~M.~Banda Guzm\'an, R.~Flores-Mendieta and J.~Hernandez,
``Baryon-meson scattering amplitude in the $1/N_c$ expansion,''
[arXiv:2305.00879 [hep-ph]].

\bibitem{cohen}
T.~D.~Cohen,
``Baryon isovector electric properties and the large $N_c$ and chiral limits,''
Phys.\ Lett.\ B \textbf{359}, 23 (1995).

\bibitem{banda1}
V.~M.~Banda Guzm\'an, R.~Flores-Mendieta, J.~Hern\'andez and F.~J.~Rosales-Aldape,
``Spin and flavor projection operators in the $SU(2N_f)$ spin-flavor group,''
Phys.\ Rev.\ D \textbf{102}, 036010 (2020).

\bibitem{part}
S.~Navas \textit{et al.} [Particle Data Group],
``Review of particle physics,''
Phys.\ Rev.\ D \textbf{110}, 030001 (2024).

\bibitem{butler2}
M.~N.~Butler, M.~J.~Savage and R.~P.~Springer,
``Strong and electromagnetic decays of the baryon decuplet,''
Nucl.\ Phys.\ B \textbf{399}, 69 (1993).

\bibitem{wu}
M.~Bertilsson and S.~Leupold,
``Goldberger-Treiman relation and Wu-type experiment in the decuplet sector,''
Phys.\ Rev.\ D \textbf{109}, 034028 (2024).

\bibitem{ash1}
J.~Ashman \textit{et al.} [European Muon Collaboration],
``A measurement of the spin asymmetry and determination of the structure function $g_1$ in deep inelastic muon-proton scattering,''
Phys.\ Lett.\ B \textbf{206}, 364 (1988).

\bibitem{ash2}
J.~Ashman \textit{et al.} [European Muon Collaboration],
``An investigation of the spin structure of the proton in deep inelastic scattering of polarized muons on polarized protons,''
Nucl.\ Phys.\ B \textbf{328},1 (1989).

\bibitem{rfm24}
R.~Flores-Mendieta, S.~A.~Garcia-Monreal, L.~R.~Ruiz-Robles and F.~A.~Torres-Bautista,
``Alternative approach to baryon masses in the $1/N_c$ expansion of QCD,''
Phys.\ Rev.\ D \textbf{109}, 114014 (2024).

\bibitem{rfm04}
R.~Flores-Mendieta,
``$V_{us}$ from hyperon semileptonic decays,''
Phys.\ Rev.\ D \textbf{70}, 114036 (2004).
[arXiv:hep-ph/0410171 [hep-ph]].

\bibitem{gk}
A.~Garc{\'\i}a and P.~Kielanowski, \textit{The Beta Decay of Hyperons}, Lecture Notes in Physics Vol.~222 (Springer-Verlag,
Berlin, 1985).

\bibitem{sha}
P.~E.~Shanahan, A.~N.~Cooke, R.~Horsley, Y.~Nakamura, P.~E.~L.~Rakow, G.~Schierholz, A.~W.~Thomas, R.~D.~Young and J.~M.~Zanotti,
``SU(3) breaking in hyperon transition vector form factors,''
Phys.\ Rev.\ D \textbf{92}, 074029 (2015).

\bibitem{heo}
Y.~Heo, C.~Kobdaj and M.~F.~M.~Lutz,
``The chiral Lagrangian with three flavors and large-$N_c$ sum rules,''
Eur.\ Phys.\ J.\ A \textbf{59}, 1 (2023).

\bibitem{krause}
A.~Krause,
``Baryon matrix elements of the vector current in chiral perturbation theory,''
Helv.\ Phys.\ Acta {\bf 63}, 3 (1990).

\bibitem{b3}
V.~Bernard, N.~Kaiser and U.~G.~Meissner,
``Chiral dynamics in nucleons and nuclei,''
Int.\ J.\ Mod.\ Phys.\ E \textbf{4}, 193 (1995).


\end{thebibliography}
\end{document}